\documentclass{article}
\usepackage[utf8]{inputenc}

\usepackage[margin=2cm]{geometry}
\usepackage[utf8]{inputenc}
\usepackage{url}
\usepackage{array}
\usepackage{amsmath,amssymb,amsfonts}
\usepackage{mathtools}
\usepackage{algorithm}
\usepackage{algpseudocode}
\usepackage{tikz}
\usepackage{graphicx}
\usepackage{subcaption}
\usepackage[export]{adjustbox}
\usepackage{bm}
\usepackage{xcolor}
\usepackage{tabularx}
\usepackage{multirow}
\usepackage{natbib}
\usepackage{amsmath}
\usepackage{amsthm}
\usepackage{amssymb}
\usepackage{amsbsy}
\usepackage{authblk} 
\usepackage{hyperref} 
\usepackage[utf8]{inputenc} 
\usepackage{siunitx} 
\usepackage{booktabs}
\usepackage{multirow}
\usepackage{ulem} 

\newcommand{\data}{\pmb{\mathcal{X}}}

\newcommand{\parametersing}{\vectGreek{\theta}}

\newcommand{\prior}{\pi}
\newcommand{\matr}[1]{\mathsf{#1}}

\newcommand{\vect}[1]{\mathbf{#1}}
\newcommand{\vectGreek}[1]{\boldsymbol{#1}}

\DeclareMathOperator{\expectation}{\mathbb{E}}
\DeclareMathOperator{\gammaDist}{Gamma}

\DeclareMathOperator{\normal}{N}
\DeclareMathOperator{\poisson}{Poisson}
\DeclareMathOperator{\prob}{\mathbb{P}}
\DeclareMathOperator{\variance}{\mathbb{V}ar}

\begin{document}\sloppy

\title{High-resolution Probabilistic Precipitation Prediction for use in Climate Simulations}
\author[1]{Sherman E. Lo}
\author[2]{Peter Watson}
\author[3]{Peter Dueben}
\author[1]{Ritabrata Dutta\thanks{Corresponding author: Ritabrata.Dutta@warwick.ac.uk}}
\affil[1]{University of Warwick, UK}
\affil[2]{University of Bristol, UK}
\affil[3]{The European Centre for Medium-Range Weather Forecasts, UK}
\date{May 2020}

\maketitle

\begin{abstract}
The accurate prediction of precipitation is important to allow for reliable warnings of flood or drought risk in a changing climate. However, to make trust-worthy predictions of precipitation, at a local scale, is one of the most difficult challenges for today's weather and climate models. This is because important features, such as individual clouds and high-resolution topography, cannot be resolved explicitly within simulations due to the significant computational cost of high-resolution simulations. 

Climate models are typically run at $\sim50-100$ km resolution which is insufficient to represent local precipitation events in satisfying detail. 
Here, we develop a method to make probabilistic precipitation predictions based on features that climate models can resolve well and that is not highly sensitive to the approximations used in individual models. To predict, we will use a temporal compound Poisson distribution dependent on the output of climate models at a location. We use the output of Earth System models at coarse resolution ($\sim\SI{50}{\kilo\metre}$) as input and train the statistical models towards precipitation observations over Wales at  $\sim\SI{10}{\kilo\metre}$ resolution. 

A Bayesian inferential scheme is provided so that the compound-Poisson model can be inferred using a Gibbs-within-Metropolis-Elliptic-Slice sampling scheme which enables us to quantify the uncertainty of our predictions. In addition, we use a Gaussian process regressor on the posterior samples of the model parameters, to infer a spatially coherent model and hence to produce spatially coherent rainfall prediction. We illustrate the prediction performance of our model by training over 5 years of the data up to 31st December 1999 and predicting precipitation for 20 years afterwards for Cardiff and Wales.
\end{abstract}

\section{Introduction and Background}
The failure to meet climate change mitigation and adaptation targets are ranked among the leading threats to global society \citep{wef2019global}. Extreme weather events caused over 500 thousand casualties and cost the economy over US\$2 trillion in damages in the past 20 years \citep{wallemacq2018economic} and these numbers are likely to increase over the coming years due to climate change. Unfortunately, there are substantial shortcomings in the ability to predict the impact of climate change on extreme events at a regional level. This is mainly caused by climate models having an insufficient resolution. State-of-the-art global climate models typically run with horizontal resolutions of $\sim$ \SIrange{50}{100}{\kilo\metre}. This is not sufficient to represent the impact of clouds or small-scale topography on local precipitation, thus making it impossible to predict local details of heavy precipitation events such as the 2013-14 UK winter floods in England and Wales \citep{schaller2016human}. 

While the resolution of global weather and climate models is steadily increasing and ``storm-resolving" simulations with a grid spacing of approximately \SI{1}{\kilo\metre} -- that allow for the explicit representation of large convective events within simulations -- have become available for research experiments \citep{Fuhrer2018,Dueben2020}, these simulations are still out of reach for operational weather and climate predictions \citep{Schulthess2019,Neumann2018}. 

Until storm-resolving simulations can be used for global predictions, the post-processing of model output from simulations at coarser resolution may allow for improvements of predictions of local precipitation. However, to make probabilistic predictions at high-resolution typically requires ensemble simulations at high-resolution (see for example \cite{Vogel2018}) which are computationally very expensive as they run several forecast trajectories to quantify predictions uncertainty. Ensemble predictions have recently also been post-processed using machine learning techniques such as random forests \citep{Hewson2020} or deep neural networks \citep{Rasp2018,Groenquist2020}. 

The \textbf{\textit{main aim}} of this paper is to develop a probabilistic methodology that allows reliable mapping of large-scale atmospheric variables from low-resolution weather or climate models to precipitation at a higher resolution. Such a tool could be useful in many applications:
\begin{itemize}
\item To improve local precipitation predictions of weather or climate models as such a tool would allow augmenting the predictions of the model with additional information from local observations and input fields 
(such as geopotential and temperature typically have a much longer prediction horizon when compared to predictions of precipitation). Therefore, the prediction horizon for our tools that use fields with long prediction horizon as inputs may be able to outperform the prediction horizon for precipitation of expensive model runs at high resolution. 
\item To generate precipitation outputs from models that do not have precipitation output fields available such as weather and climate models based on deep learning \citep{Dueben2018,Scher2019,Weyn2019,Rasp2020}.
\item To generate probabilistic predictions for precipitation from deterministic simulations. 
\end{itemize}

To achieve this, we use six coarse-resolution meteorological variables, otherwise known as model fields, from the ERA5 reanalysis dataset \citep{ERA5}. In the reanalysis dataset, observations are continuously assimilated to a global atmospheric model to generate atmospheric states as available in weather and climate models but consistent with the local weather observations at a given space and time. This data is used as the input feature vector to our model to predict the high-resolution precipitation observations that are available from the E-obs dataset \citep{E_obs} over Wales, chosen due to its complex topography, making local precipitation prediction challenging. The focus on a region of the size of Wales also represents a good compromise between size and computational cost.

ERA5 does provide precipitation as a variable which may yield further predictive information by using it as an additional input variable. However, we chose not to use precipitation as an input variable because we aim to provide a generic model which can be used by different dynamical weather and climate models which may not provide precipitation as a variable or which may represent precipitation quite differently within simulations.

A parametric distribution is considered here as the distribution of rainfall at each location and time point, namely, the compound Poisson distribution which was previously used for modelling the occurrence and quantity of precipitation in \cite{revfeim1984initial} and \cite{dunn2004occurrence}. 
For precipitation modelling in \cite{holsclaw2017bayesian} and \cite{bertolacci2019climate}, the temporal dependency of rainfall has been modelled by a hidden inhomogeneous Markov model with an emission distribution consisting of a mixture of zero mass and finite mixtures of Gamma distributions. The compound Poisson distribution is a mixture of zero mass and an infinite number of Gamma distributions, thus an extension to the finite mixture. We expect that this approach will perform better when modelling high precipitation.

Further, we assume that the parameters of the compound Poisson distribution are dependent on the input variables at each location, meaning the parameters of the distribution of rainfall at a day and a location will be determined by a function of the time-series of the input variables at that location. 
In \cite{garcia1995analysis, dzupire2018poisson}, the temporal structure of precipitation was modelled by making the parameters of the compound Poisson sinusoidal functions of time to capture seasonality behaviour in Spain and Malawi. Similarly, to impose both the time dependency and influence of the input variables on the model parameters, here we consider a
latent auto-regressive moving average (ARMA) process for each of the parameters of the compound Poisson model dependent on the input variables and the parameters and rainfall at previous and present time points, which are linked to the parameters via an exponential link function.
The details of our model are described in Section~\ref{sec:model_prior}. 
To train our model, in Section~\ref{sec:computing}, we devise a Bayesian inferential framework using a Markov chain Monte Carlo (MCMC) scheme targeting the posterior distribution of the model parameters given the dataset, along with the prior distributions used. In Section~\ref{sec:cpmcmcgp}, we use a Gaussian process regressor \citep{williams1996gaussian} on the posterior samples of model parameter to learn a spatially coherent model, which can produce median rainfall prediction in a spatially coherent manner.

We illustrate the performance of our model by training on 5 years of data up to 31st December 1999, inclusive, and using the trained model to predict the precipitation between 1st January 2000 to 31st July 2019, inclusive, first for a single location close to Cardiff to validate our model and then for Wales, applying the temporal model to each location on the high-resolution grid independently. We compare our precipitation prediction with the representation of precipitation in short-term predictions with the Integrated Forecasting System (IFS). IFS is the underlying global weather model that is used to generate ERA5 and is based on the full three-dimensional state of the atmosphere. As a result, it should be noted that IFS is a high-quality and competitive benchmark.
Using only six meteorological variables, our mapping produces local predictions of precipitation that have skill comparable to the representation of precipitation within the IFS, which is at a higher resolution.

Furthermore, our methodology provides a way of producing probabilistic predictions of precipitation at a high-resolution which agrees well with observed precipitation data. 
We have used the MCMC samples 
from the posterior distribution of the model parameters given the observed data to produce probabilistic forecasts of precipitation, hence quantifying the uncertainty of our prediction. The dataset, our competitive benchmark, IFS, and our tuning parameters are described in Section~\ref{sec:dataset_benchmark_tuning}. Results in Section~\ref{sec:results} show that our method is competitive with IFS. In addition, we illustrate the uncertainty quantification of our precipitation forecasts. The conclusion is in Section~\ref{sec:conclusion}.

\section{Model and Methods}
\subsection{Statistical Model}
\label{sec:model_prior}
On day $t = 1, 2, \ldots, T$ and location $s=1,2,\ldots, S$, we will denote the observed precipitation as $Y_{t,s}$ and $\vect{x}_{t,s}=\left(x^1_{t,s},x^2_{t,s},\ldots,x^R_{t,s}\right)$ will be the vector containing $R$ standardised (mean 0 and standard deviation 1 across space and time) input variables. The input variables are derived from meteorological fields, generated by coarse-resolution weather and climate models, which were interpolated to the $S$ locations on the high-resolution grid by using a bivariate spline approximation on the rectangular mesh. For a given location $s$, the time series of the input vectors and observed precipitation until day $T$ will be denoted by  $\matr{X}^T_s=(\vect{x}_{1,s},\vect{x}_{2,s},\ldots ,\vect{x}_{T,s})$ and $\matr{Y}^T_s=(Y_{1,s},Y_{2,s},\ldots ,Y_{T,s})$ respectively. Together, for all $S$ locations, $\mathcal{X}^T=\{\matr{X}^T_1,\matr{X}^T_2,\ldots,\matr{X}^T_S\}$ and $\mathcal{Y}^T=\{\matr{Y}^T_1,\matr{Y}^T_2,\ldots,\matr{Y}^T_S\}$ are the collection of all of the time-series until day $T$ of input vectors and observed precipitation respectively.

We further consider $Z_{t,s}$, the occurrence of precipitation at the $t$-th day and location $s$, as a latent variable such that the occurrence was not available in the training data. The distribution of $Z_{t,s}$ and the conditional distribution of $Y_{t,s}|Z_{t,s}$ are assumed to be
\begin{align}
  Z_{t,s}&\sim\poisson(\lambda_{t,s})
  \\
  Y_{t,s}|Z_{t,s}&\sim\gammaDist\left(\dfrac{Z_{t,s}}{\omega_{t,s}},\dfrac{1}{\omega_{t,s}\mu_{t,s}}\right)
\end{align}
where  $\lambda_{t,s}$, $\mu_{t,s}$ and $\omega_{t,s}$ are time and location specific parameters. The parameterisation of $Y_{t,s}|Z_{t,s}$ was chosen such that $\expectation[Y_{t,s}|Z_{t,s}]=Z_{t,s}\mu_{t,s}$ and $\variance[Y_{t,s}|Z_{t,s}]=Z_{t,s}\omega_{t,s}\mu_{t,s}^2$ and the marginal distribution of $Y_{t,s}$ is known to be the compound Poisson distribution \citep{dunn2004occurrence}. $Z_{t,s} = 0$ and $Z_{t,s} > 0$ correspondingly indicate a dry day and a wet day, making the distribution of $Y_{t,s}$ a mixture of  point  mass  at  0  and  a  continuous  distribution for rainfall on wet days. 

\begin{figure}
    \centering
    \includegraphics[width = .5\textwidth]{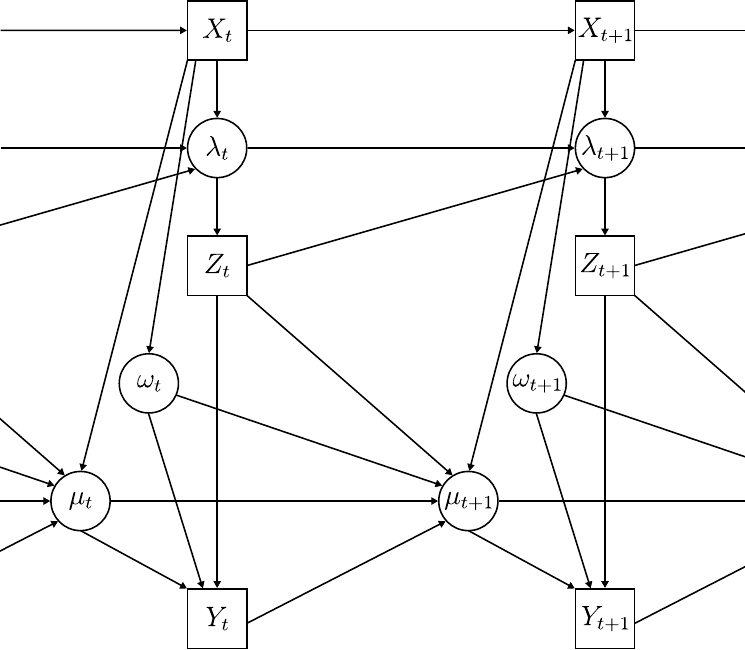}
    \caption{\textbf{Dependency structure in our model} of the parameters of the compound Poisson distribution on the input variables, latent states and their values at present and past time points. $X_t$, $Z_t$ and $Y_t$ are correspondingly the input variables, number of times it rained and total amount of rain on $t$-th day. The parameters of the compound Poisson distribution of rainfall on day $t$ is denoted by $\lambda_t$, $\mu_t$ and $\omega_t$. }
    \label{fig:model}
\end{figure}

To impose temporal dependency on the model parameters $\lambda_{t,s}$, $\mu_{t,s}$ and $\omega_{t,s}$ at day $t$ and location $s$ and their dependency on the input variables, we consider a latent moving-average and auto-regressive process dependent on 
the parameters, latent states and the input variables at the present and previous time step. These processes are linked to the model parameters through an exponential link function, for each of the parameters of the model.  
The latent ARMA processes, their links to the parameters $\lambda_{t,s}$, $\mu_{t,s}$ and $\omega_{t,s}$ and their dependency on the input variables and
their values at previous time-points as well as previous realisations of $Z_{t,s}$ and $Y_{t,s}$ can be described as,
\begin{align}
  \lambda_{t,s} &= \exp\left[\vectGreek{\beta}^{\lambda}_{s} \cdot \vect{x}_{t,s} + \Phi^{\lambda}_{t,s} + \Gamma^{\lambda}_{t,s} + k^{\lambda}_{s}\right]
  \\
  \mu_{t,s} &= \exp\left[\vectGreek{\beta}^{\mu}_{s} \cdot \vect{x}_{t,s} + \Phi^{\mu}_{t,s} + \Gamma^{\mu}_{t,s} + k^{\mu}_{s}\right]
  \\
  \omega_{t,s} &= \exp\left[\vectGreek{\beta}^{\omega}_{s} \cdot \vect{x}_{t,s} + k^{\omega}_{s}\right]
\end{align}
where $\vectGreek{\beta}^{\lambda}_{s},\vectGreek{\beta}^{\mu}_{s},\vectGreek{\beta}^{\omega}_{s}$ are the location-specific regression parameters, $k^{\lambda}_{s}, k^{\mu}_{s}, k^{\omega}_{s}$ are location-specific constants, $\Phi^{\lambda}_{t,s}, \Phi^{\mu}_{t,s}$ are the autoregressive (AR) components and $\Gamma^{\lambda}_{t,s}, \Gamma^{\mu}_{t,s}$ are the moving average (MA) components at time $t$ and location $s$. 
The AR components with lag $p$ are
\begin{align}
  \Phi^{\lambda}_{t,s} &= \sum_{i=1:t>i}^p \phi^{\lambda}_{i,s}\left[
      \ln\lambda_{t-i,s}-k^{\lambda}_{s}
  \right]
  \\
   \Phi^{\mu}_{t,s} &= \sum_{i=1:t>i}^p \phi^{\mu}_{i,s} \left[
      \ln\mu_{t-i,s}-k^{\mu}_{s}
  \right]
\end{align}
and the MA components with lag $q$ are
\begin{align}
  \Gamma^{\lambda}_{t,s} &= \sum_{i=1:t>i}^q
       \gamma^{\lambda}_{i,s}\dfrac{Z_{t-i,s}-\lambda_{t-i,s}}{\sqrt{\lambda_{t-i,s}}}
  \\
  \Gamma^{\mu}_{t,s} &= \sum_{i=1:t>i}^q
      \gamma^{\mu}_{i,s}\dfrac{y_{t-i,s}-Z_{t-i,s}\mu_{t-i,s}}{\mu_{t-i,s}\sqrt{Z_{t-i,s}\omega_{t-i,s}}}
  \ .
\end{align}
The terms $\phi^{\lambda}_{i,s},\phi^{\mu}_{i,s}\gamma^{\lambda}_{j,s},\gamma^{\mu}_{j,s}$ are ARMA parameters for $i=1,2,\ldots,p$ and $j=1,2,\ldots,q$ at location $s$. The motivation behind this emulation is that the AR process depends on past parameters whereas the MA process also depends on past realisations of $Y_{t,s}$ and $Z_{t,s}$. The denominator in the MA components help to stabilise the process by using the variance, $\variance[Z_{t,s}]=\lambda_{t,s}$ and $\variance[Y_{t,s}|Z_{t,s}]=Z_{t,s}\omega_{t,s}\mu_{t,s}^2$. Our proposed time-series model and the dependency structure of the parameters are illustrated in Figure~\ref{fig:model}. The model at location $s$, is parameterised by 

\begin{equation}
\label{eq:param_sing_loc}
    \parametersing_s=\left(k^{\lambda}_{s}, k^{\mu}_{s}, k^{\omega}_{s}, \vectGreek{\beta}^{\lambda}_{s}, \vectGreek{\beta}^{\mu}_{s}, \vectGreek{\beta}^{\omega}_{s}, \vectGreek{\phi}^{\lambda}_{s}, \vectGreek{\phi}^{\mu}_{s}, \vectGreek{\gamma}^{\lambda}_{s}, \vectGreek{\gamma}^{\mu}_{s} \right)
\end{equation}
where $\vectGreek{\phi}^{\lambda}=(\phi^{\lambda}_{1,s}, \phi^{\lambda}_{2,s},\ldots, \phi^{\lambda}_{p,s}) $,
$\vectGreek{\phi}^{\mu}_{s}=(\phi^{\mu}_{1,s}, \phi^{\mu}_{2,s},\ldots, \phi^{\mu}_{p,s}) $, 
$\vectGreek{\gamma}^{\lambda}_{s}=(\theta^{\lambda}_{1,s}, \gamma^{\lambda}_{2,s},\ldots, \gamma^{\lambda}_{q,s}) $,     $\vectGreek{\gamma}^{\mu}_{s}=(\gamma^{\mu}_{1,s}, \gamma^{\mu}_{2,s},\ldots, \gamma^{\mu}_{q,s}) $ and $\parametersing_s$ is a $L$-dimensional ($L=3+3R+2p+2q$) parameter. 

\subsection{Bayesian Inference}
\label{sec:computing}
The inference of the above model from data is equivalent to learning the $L$-dimensional parameter in Equation~\ref{eq:param_sing_loc} for each location. 
We can quantify the  uncertainty of the model parameter $\parametersing_s$, by its posterior distribution $\prior(\parametersing_s|\data_s)$ given the observed dataset $\data_s = \left(\matr{X}^T_s, \matr{Y}^T_s \right)$ until day $T$ correspondingly for a location $s$. The posterior distribution of the parameter $\parametersing_s$ is obtained by Bayes' theorem as
$$\prior(\parametersing_s|\data_s) = \frac{\prior(\parametersing_s)
 p(\data_s|\parametersing_s)}{m(\data_s)},
$$
where $\prior(\parametersing_s)$, $p(\data_s|\parametersing_s)$ and $m(\data_s) = \int\prior(\parametersing_s)p(\data_s|\parametersing_s)d\parametersing_s$ are, correspondingly, the prior distribution on the parameter $\parametersing_s$, the likelihood function, and the marginal likelihood.
If the likelihood function could be evaluated, at least up to a normalizing constant, then the posterior distribution could be approximated by drawing a representative sample of parameter values from it using (Markov chain) Monte Carlo sampling schemes \citep{Robert_2005}. In the following, we propose a Markov chain Monte Carlo algorithm using adaptive Metropolis-Hastings \citep{metropolis1953equation, hastings1970monte, haario2001adaptive, roberts2009examples}, slice sampling \citep{neal2003slice} and elliptical slice sampling \citep{murray2010elliptical} within a Gibbs sampling \citep{geman1984stochastic} scheme of which the invariant distribution is the posterior distribution of the parameters of our model, enabling us to draw samples from the posterior distribution.

\paragraph{Prior and posterior distribution:} For each location $s\in\{1,2,\ldots,S\}$, the following prior distributions were used
\begin{equation}
\label{eq:prior_sing}
    \begin{pmatrix}
        k^{\lambda}_{s} \\ k^{\mu}_{s} \\ k^{\omega}_{s} \\ \vectGreek{\beta}^{\lambda}_{s} \\ \vectGreek{\beta}^{\mu}_{s} \\ \vectGreek{\beta}^{\omega}_{s}
    \end{pmatrix}
    \sim\normal\left(
        \begin{pmatrix}
            k^{\lambda}_0 \\ k^{\mu}_0 \\ k^{\omega}_0 \\ \vect{0} \\ \vect{0} \\ \vect{0}
        \end{pmatrix}
        ,\left(\tau^{\beta}_{s}\right)^{-1} I
    \right)
\end{equation}
\begin{equation}
    \begin{pmatrix}
        \vectGreek{\phi}^{\lambda}_{s} \\
        \vectGreek{\phi}^{\mu}_{s} \\
        \vectGreek{\gamma}^{\lambda}_{s} \\
        \vectGreek{\gamma}^{\mu}_{s} \\
    \end{pmatrix}
    \sim\normal\left(
        \vect{0}, \left(\tau^{\text{ARMA}}_{s}\right)^{-1} I
    \right)
\end{equation}
where the hyper-parameters are $k^{\lambda}_0$, $k^{\mu}_0$, $k^{\omega}_0$, $\tau^{\beta}_{s}$ and $\tau^{\text{ARMA}}_{s}$ and $I$ is the identity matrix. The hyper-parameters $k^{\lambda}_0$, $k^{\mu}_0$, $k^{\omega}_0$ were chosen to be $-0.46, 1.44$ and $-0.45$ correspondingly using the naive estimator in  Equation~\ref{eq:initialestimate} below for the British Isles. For $\vectGreek{\tau}_s=\left(\tau^{\beta}_{s}, \tau^{\text{ARMA}}_{s}\right)$, the priors
\begin{equation}
\label{eq:prior_tau}
    \tau^{\beta}_{s} \sim \gammaDist(2.8, 2.3^{-1}) \  , \ 
    \tau^{\text{ARMA}}_{s} - 16 \sim \gammaDist(1.3, 65^{-1})
\end{equation}
were used. The hidden variables $\matr{Z}^T_s = (Z_{1,s},Z_{2,s},\ldots,Z_{T,s})$ have discrete uniform distribution over non-negative integers and independent for all the locations. For all the other priors described above, they are also considered to be independent for all locations $s=1,2,\ldots,S$.

After observing the precipitation data $\matr{Y}^T_s$ and feature vectors $\matr{X}^T_s$, the posterior distribution of the parameters of interest are
\begin{equation}
\label{eq:post_sing_loc}
    \pi(\matr{Z}^T_1,\matr{Z}^T_2,\ldots,\matr{Z}^T_S,\parametersing_1,\parametersing_2,\ldots,\parametersing_S, \vectGreek{\tau}_1,\vectGreek{\tau}_2,\ldots,\vectGreek{\tau}_S|\data)
    \propto
    \prod_{s=1}^S
    p(\matr{Y}^T_s |\parametersing_s,\matr{Z}^T_s,\matr{X}^T_s)p(\matr{Z}^T_s|\parametersing_s,\matr{X}^T_s)\pi(\parametersing_s|\vectGreek{\tau}_s)\pi(\vectGreek{\tau}_s)
    \ ,
\end{equation}
where $\matr{Z}^T_s = (Z_{1,s},Z_{2,s},\ldots,Z_{T,s})$ are hidden variables and $\parametersing_s=\left(k_{\lambda,s}, k_{\mu,s}, k_{\omega,s}, \vectGreek{\beta}_{\lambda,s}, \vectGreek{\beta}_{\mu,s}, \vectGreek{\beta}_{\omega,s}, \vectGreek{\phi}_{\lambda,s}, \vectGreek{\phi}_{\mu,s}, \vectGreek{\gamma}_{\lambda,s}, \vectGreek{\gamma}_{\mu,s} \right)$ at location $s$.

\paragraph{MCMC sampling:} For a given location $s$, we propose a random scan Gibbs sampling scheme \citep{geman1984stochastic} to sample from the posterior distribution in Equation~\ref{eq:post_sing_loc}. Random Scan Gibbs sampling scheme is an iterative scheme. At the $i$-th iteration, we update one of the components of $(\parametersing_s, \vectGreek{\tau}_s, \matr{Z}^T_s)$ correspondingly with probability $(p_{\theta}, p_{\tau}, p_{Z})$ by sampling from their conditional distribution given the values of the other components at the $(i-1)$-th step. We choose $(p_{\theta}, p_{\tau}, p_{Z})$ proportional to the dimension of each of the components, such that a component with larger dimensions will get updated more frequently than a component with smaller dimension. At any of the $i$-th iteration, the update of the chosen component can be done as follows:
\begin{itemize}
    \item Update $\parametersing_s$ at the $i$-th iteration by sampling $\parametersing_s^{(i)}$ from $\pi(\parametersing_s|\matr{X}^T_s, \matr{Y}^T_s, \matr{Z}^{T,(i-1)}_s,\vectGreek{\tau}_s^{(i-1)})$, using elliptical slice sampling \citep{murray2010elliptical}. Loosely, elliptical slice sampling proposes a sample from a Gaussian prior distribution. It uses a weighted average of the proposed sample with the current MCMC sample to search for the next Markov step which satisfies the slice sampling scheme \citep{neal2003slice}. The advantage of elliptical slice sampling is that it does not have any tuning parameters unlike other MCMC methods such as Hamiltonian Monte Carlo \citep{neal2011mcmc, hoffman2014no}. 
    \item Update $\vectGreek{\tau}_s$ at the $i$-th iteration by sampling $\vectGreek{\tau}^{(i)}_s$ from $\pi(\vectGreek{\tau}_s|\matr{X}^T_s, \matr{Y}^T_s,\matr{Z}^{T,(i-1)}_s,\parametersing^{(i-1)}_s)$, using adaptive Metropolis Hastings \citep{metropolis1953equation, hastings1970monte, haario2001adaptive}, but specifically the version presented in \cite{roberts2009examples}.
    \item Update $\matr{Z}^T_s$ at the $i$-th iteration by sampling $\matr{Z}^{T,(i)}_s$ from $p(\matr{Z}^T_s|\matr{X}^T_s, \matr{Y}^T_s, \parametersing^{(i-1)}_s,\vectGreek{\tau}^{(i-1)}_s)$ by selecting, at random, a time point $t$ and then sampling $p(Z_{t,s}|\matr{Z}^{T,(i-1)}_{-t,s},\matr{X}^T_s,\matr{Y}^T_s, \parametersing^{(i-1)}_s,\vectGreek{\tau}^{(i-1)}_s)$ using slice sampling \citep{neal2003slice} where $\matr{Z}^{T,(i-1)}_{-t,s}$ is a vector containing all elements of $\matr{Z}^T_s$ after $(i-1)$-th iteration except for the $t$-th time point. In other words, one random time point is sampled. It was found the typical values of $Z_{t,s}$ take integer values from zero to about 6 so the search space in slice sampling is quite small, making it quite effective for this particular example.
\end{itemize}
The iteration is run until $\mathbf{n}$ many steps, of which the first $\mathbf{n}_{\text{burn-in}}$ steps are discarded based on the trace plots of the Markov chain \citep{geyer1992practical} to make sure the samples are drawn only after the Markov chain has converged to the invariant distribution, in our problem the posterior distribution in Equation~\ref{eq:post_sing_loc}. 

The initial values for the Markov chain Monte Carlo (MCMC) were set using the prior mean for $k^{\lambda}_{s}, k^{\mu}_{s}, k^{\omega}_{s}$ and $\matr{Z}^T_s$. $k^{\lambda}_{s}, k^{\mu}_{s}, k^{\omega}_{s}$ were initialised using naive method of moments estimator for a i.i.d.~sample of $\vect{Y}_s$, or a slightly modified version,
which are
\begin{align}
\nonumber
\exp\left[\widehat{k}^{\lambda}_{s}\right]&=\widehat{\lambda}_s=\ln\left[
    \dfrac{T+1/2}{T-\sum_{t=1}^T\mathbb{I}(y_{t,s}>0)+1}
\right]
\\
\nonumber
\exp\left[\widehat{k}^{\mu}_{s}\right] &= \widehat{\mu}_s = \dfrac{
   \overline{y}_s    
}
{
    \widehat{\lambda}_s
}
,
\\
\exp\left[\widehat{k}^{\omega}_{s}\right] &= \widehat{\omega}_s = \dfrac{
   \sigma_{y,s}^2
}
{
    \widehat{\lambda}_s\widehat{\mu}_s^2
}
-1
\label{eq:initialestimate}
\end{align}
where $\overline{y}_s$ and $\sigma_{y,s}^2$ are the sample mean and variance. It was found that for the given dataset of Wales, these estimators are real for large $T$, for example, longer than a year. $\matr{Z}^T_s$ was initialised by evaluation the conditional expectation using
\begin{equation}
\expectation\left[Z_{t,s}|Y_{t,s}=y_{t,s},\parametersing_s,\matr{X}_s^T\right] =
  \begin{cases}
  0 & \text{ if } y_{t,s} = 0 \\ 
  \dfrac{\sum_{z=1}^{\infty}z \prob(Z_{t,s}=z|Y_{t,s}=y_{t,s},\parametersing_s,\matr{X}_s^T)}{\sum_{z=1}^{\infty}\prob(Z_{t,s}=z|Y_{t,s}=y_{t,s},\parametersing_s,\matr{X}_s^T)} & \text{ if } y_{t,s}>0 
  \end{cases}
\end{equation}
and then rounded to the nearest non-zero integer for $t=1,2,\cdots,T$. The sum can be truncated to only evaluate terms which are large enough to contribute to a sum \citep{dunn2005series, lo2020characterisation}.

As we do not consider the spatial dependence between different locations, the posterior distributions at each location are independent and a random Gibbs sampling scheme targeting each of these location-specific posterior distributions can be run in parallel asynchronously and independently.

\paragraph{Posterior predictive distribution of precipitation}
In previous sections, we provide details of how to draw samples from the posterior distribution described in Equation~\ref{eq:post_sing_loc} given $\data_s$ at time $T$. 
If we have drawn $M$ samples from the posterior distribution, we can use each of those parameter values and input vectors available at a time point $t>T$ with our model to provide a prediction of precipitation of $Y_s^t$ at a location $s$ and time $t>T$. 
Hence, using $M$ samples from the posterior distribution we will get $M$ predictions of precipitations, $\lbrace Y_{s,1}^t, \ldots, Y_{s,M}^t  \rbrace$, marginalising the parameters which provides samples from the posterior predictive  distribution $p\left(Y_{s}^t|\data_s,\matr{X}^t_s\right)$ of precipitation at location $s$ and time $t>T$ given observed precipitation data until time $T$. This model and the corresponding predictions will be later referred as CP-MCMC.

Furthermore, each prediction, which is a time series for post-processed precipitation at each location, is using a sample from the posterior distribution and can be done independently 
and can therefore be parallelised independently and asynchronously.

\subsection{Spatially coherent model using Gaussian process regression}
\label{sec:cpmcmcgp}
An important goal in meteorology is to produce probabilistic rainfall predictions at high-resolution ($\sim$10km) that have a realistic spatial structure and give well-calibrated uncertainty, such that the spatial correlations of predicted rainfall over a large area approximately match to the observed rainfall. 
Having coherent rainfall prediction over a large area is a crucial pre-requisite for hydrological and flood risk modelling. Recent reviews by \cite{maraun2019statistical, gutierrez2019intercomparison}, highlight that getting spatial correlations right is a particular challenge, which none of the existing methods evaluated there could do well. In this context, we notice that our model in Section~\ref{sec:model_prior} has been only inferred at different locations independently and hence they do not have any spatial coherence. 

To improve spatial coherence, we assume that the model parameter $\parametersing_s = \parametersing(w_s)$ is a $L$-dimensional smooth function of spatial topology $w = (\mbox{latitude, longitude})$ at each location and we would like to learn spatially coherent values of model parameters $\left(\theta^{'}_i(w_{1}), \theta^{'}_i(w_{2}), \ldots, \theta^{'}_i(w_{S})\right)$ for $i=1,\ldots,L$ and at all the $S$ locations. We assume independently each of the functions $\theta_i(w_s), \ i = 1, \ldots, L$ corresponding to the $L$ dimensions of $\parametersing(w_s)$ follows a Gaussian process \citep{williams1996gaussian}. The Gaussian process on the function at dimension $i \in L$ imposes the following multivariate Gaussian distribituion on the values of the function at all the $S$ locations on the high resolution grid: 
\begin{equation}
\label{eq:Gp}
\left(\theta_i(w_{1}), \theta_i(w_{2}), \ldots, \theta_i(w_{S})\right)  \sim N(m_i(w_{1}), m_i(w_{2}), \ldots, m_i(w_{S}), K),
\end{equation}
where $m_i(w)$ is the mean function and $K$ is the $S\times S$ dimensional kernel matrix with $(i,j)$-th element being $K_{i,j} = K(w_i,w_j) = \sigma_{\theta}^2\exp{\left( -\frac{1}{2l^2}||w_i-w_j||^2 \right)}$. We consider $$\left(\theta^{(1)}_i(w_{j}), \theta^{(2)}_i(w_{j}), \ldots, \theta^{(M)}_i(w_{j})\right),$$ $M$ samples from the posterior distribution $(\theta_i(w_{j})|\data)$ of the $i$-the parameter at $j$-th location as drawn in Section~\ref{sec:computing}, being identical and independently distributed samples from $ N(\theta_i(w_{j}), \sigma^2_e)$, where $\sigma^2_e$ is the noise variance. This assumption can be justified asymptotically via Bernstein-von-Misses theorem \citep{lecam1953some}. Under the spatial smoothness assumptions in Equation~\ref{eq:Gp}, we learn $(m(w_1),\ldots,m(w_S), \sigma^2_f, \sigma^2_{\theta}, l)$ for each of the $L$ parameters using maximum likelihood estimation scheme, which provides us a distribution on the values of function $\parametersing(w_s), \ s=1, \ldots, S$. Each independent and identical sample from this distribution provides a value for the spatially coherent model parameters $\left(\theta^{'}_i(w_{1}), \theta^{'}_i(w_{2}), \ldots, \theta^{'}_i(w_{S})\right)$ for $i=1,\ldots,L$ at all the $S$ locations, which we will use for spatially coherent prediction of rainfall in future for all of the Wales. This model and the corresponding predictions will be later referred as CP-MCMC-GP.
\section{Dataset, Benchmark and Tuning parameters}
\label{sec:dataset_benchmark_tuning}

\paragraph{Dataset:} We use a selection of model fields, which can be resolved reasonably well by typical climate models, from the ERA5 reanalysis dataset \citep{ERA5} as input variables for our model. 
The reanalysis data is based on the IFS, a numerical weather prediction system which solves the physical equations of atmospheric motion that is used for operational weather predictions at the European Centre for Medium-Range Weather Forecasts (ECMWF). To generate ERA5, observations are assimilated in 12-hour windows into the IFS model to yield the best estimate of the state of the atmosphere during each point in time. We take these historical weather state estimates and use them to predict the high-resolution precipitation observations, made available in the E-obs dataset \citep{E_obs}, recorded daily on land at a resolution of \ang{0.1} longitude and latitude. 
Both of the datasets were retrieved for Wales between 1st January 1979 and 31st July 2019 inclusive. 

The input fields were obtained four times per day at a resolution of $\ang[parse-numbers=false]{{5}/{9}}$ longitude and $\ang[parse-numbers=false]{{5}/{6}}$ latitude. This is the level of resolution used in the climate model configuration for the UK climate prediction 2018 global simulations \cite{Murphy2018} (the grid spacing is approximately \SI{50}{\kilo\metre} over Wales). The model fields include the following variables: total column liquid water content, geopotential at \SI{500}{\hecto\pascal}, temperature at \SI{850}{\hecto\pascal}, eastward and northward components of wind velocity $u$ and $v$ at \SI{850}{\hecto\pascal} and specific humidity at \SI{850}{\hecto\pascal}. Based on these fields, we have derived wind speed and two more custom fields and used them as additional inputs. The latter two combine information from water content and advection and are defined as $\sqrt{(u\frac{\partial A}{\partial x})^2 + (v\frac{\partial A}{\partial y})^2}$ with $A$ being either total column liquid water content or the specific humidity and calculated numerically using finite central difference for every time point (4 times a day). These custom fields were included as inputs to our model 
as a weak linear correlation with the precipitation was found in our exploratory analysis.

The input vectors were obtained by using daily means of the smoothed model fields which were then interpolated to the same high-resolution grid of observed rainfall as our statistical model requires the input variables to have the same resolution as the output. 
The smoothed model fields (available at about $\ang[parse-numbers=false]{{5}/{9}}$ longitude and $\ang[parse-numbers=false]{{5}/{6}}$ latitude) were interpolated to the high-resolution grid (\ang{0.1} longitude and latitude), at which the precipitation was recorded, using a bivariate spline interpolation scheme on a rectangular mesh. The input fields were standardised across space and time to have mean 0 and standard deviation 1, giving the input vectors required for our model.

\paragraph{Benchmark:}For our benchmark, we use precipitation predictions available from short-term forecast simulations with IFS as they were performed to provide the background state of the 4DVar data-assimilation when ERA5 was generated. 
There are two forecast simulations started each day at 6 am and 6 pm. We extract the precipitation fields for the first 12 hours of each simulation to reproduce daily precipitation - this is presently the optimal way to derive meaningful precipitation predictions from a dynamical model that is consistent with the large-scale fields in the ERA5 reanalysis data. Note that the IFS precipitation predictions provide only a deterministic prediction for each point in space and time, which requires simulating the full three-dimensional state of the global atmosphere, and are computationally expensive. In contrast, our mapping provides a probabilistic prediction, requires only 6 local model fields as input, and can be generated at a much lower cost once the model is trained. Furthermore, the precipitation data was mapped directly from the native Gaussian grid of ERA5 (N320) to the high-resolution grid (\ang{0.1} longitude and latitude) of the observations and was not mapped to the coarse resolution grid at $\ang[parse-numbers=false]{{5}/{9}}$ longitude and $\ang[parse-numbers=false]{{5}/{6}}$ latitude first. The IFS precipitation predictions have therefore a higher spatial resolution ($\sim\SI{30}{\kilo\metre}$) when compared to the model fields that serve as input to our model, making it an especially challenging benchmark.

\paragraph{Tuning parameters of model:}The number of ARMA terms in our model was selected to be $p=q=5$ because it was found that the sample partial autocorrelation has significant terms up to lag 5. Furthermore, larger $p$ and $q$ were avoided to allow the number of model field terms to contribute more to the model than the ARMA terms. 
In the random Gibbs sampling scheme, $(p_{\theta}, p_{\tau}, p_{Z})$ were chosen to be proportional to $1$, $0.2$ and $T \times 3\times 10^{-3}$, reflecting the dimension of each of the components.
For the inference of the posterior distribution of the single location model at Cardiff, we investigated different training set sizes. After $\mathbf{n}=50,000$ Gibbs steps, for the inference using 1 year, 5 years, 10 years and 20 years worth of training data we discard $\mathbf{n}_{\text{burn-in}}=$ 10,000, 15,000, 30,000 and 40,000 respectively many initial samples, based on a trace plot of the MCMC chain \citep{geyer1992practical}. For the multiple locations model in Wales, we used $\mathbf{n}=20,000$ Gibbs steps with a burn-in of $\mathbf{n}_{\text{burn-in}}=15,000$. For the training of CP-MCMC-GP, we use $M=100$ posterior samples. 

\section{Results and Discussion}
\label{sec:results}
\subsection{Model validation at a single location: Cardiff}
\label{sec:cardiff_result}
\begin{figure}[ht!]
\centering
        \begin{subfigure}{0.49\textwidth}
         \includegraphics[width= \textwidth]{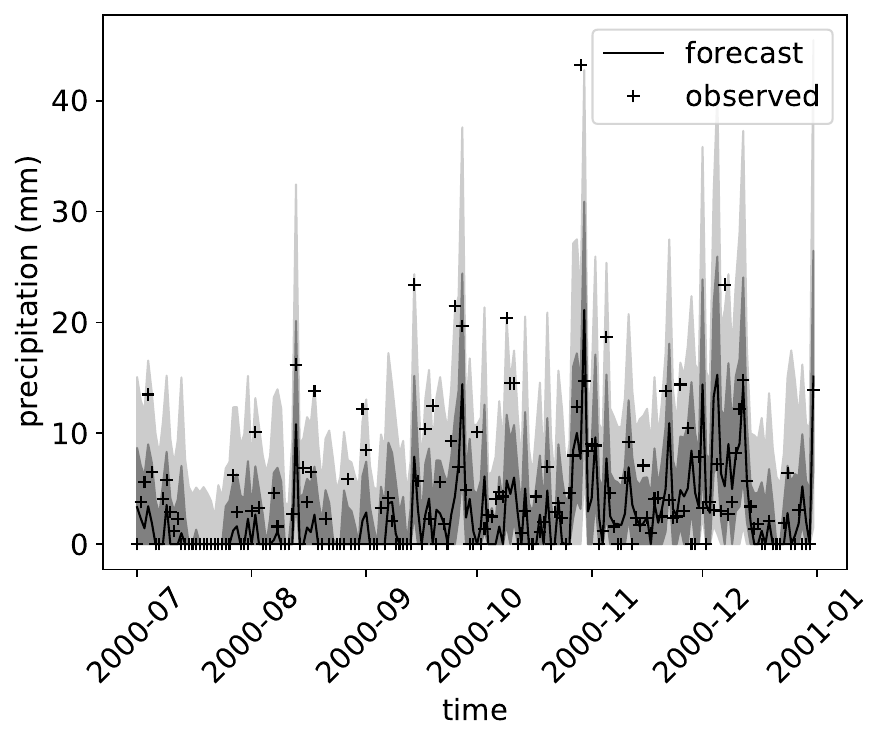}
    \caption{Year: 2000}
    \end{subfigure}
    \begin{subfigure}{0.49\textwidth}
         \includegraphics[width= \textwidth]{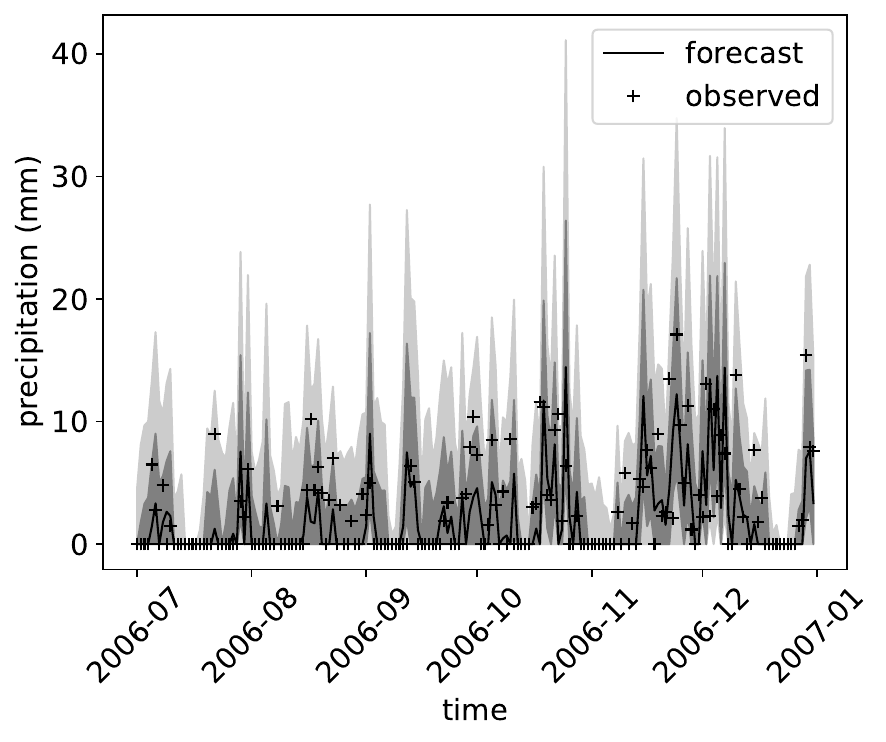}
    \caption{Year: 2006}
    \end{subfigure}
        \begin{subfigure}{0.49\textwidth}
         \includegraphics[width= \textwidth]{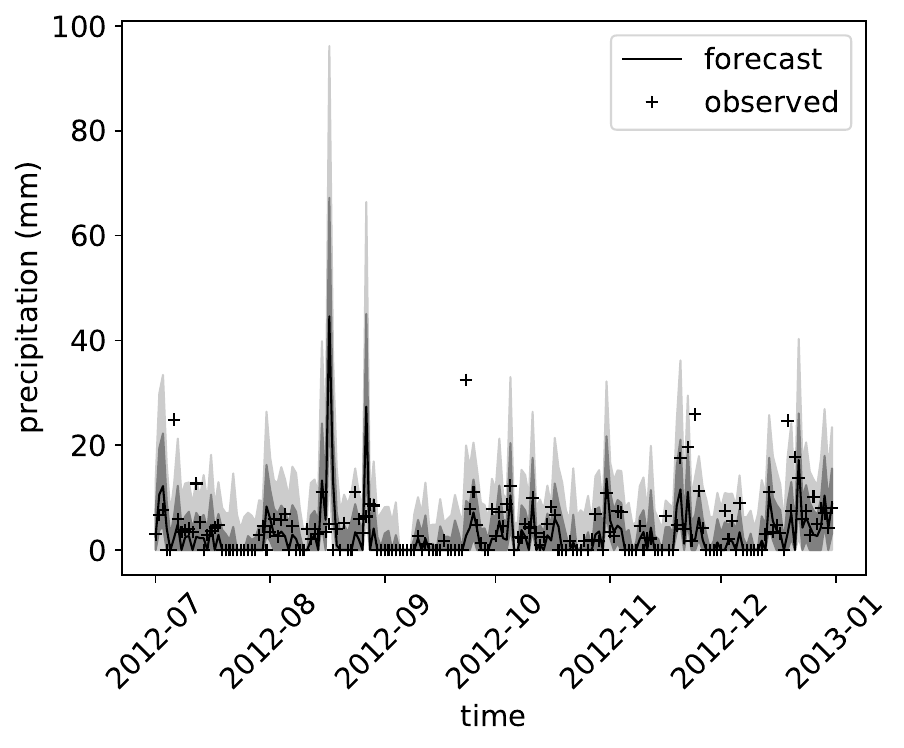}
    \caption{Year: 2012}
    \end{subfigure}
            \begin{subfigure}{0.49\textwidth}
         \includegraphics[width= \textwidth]{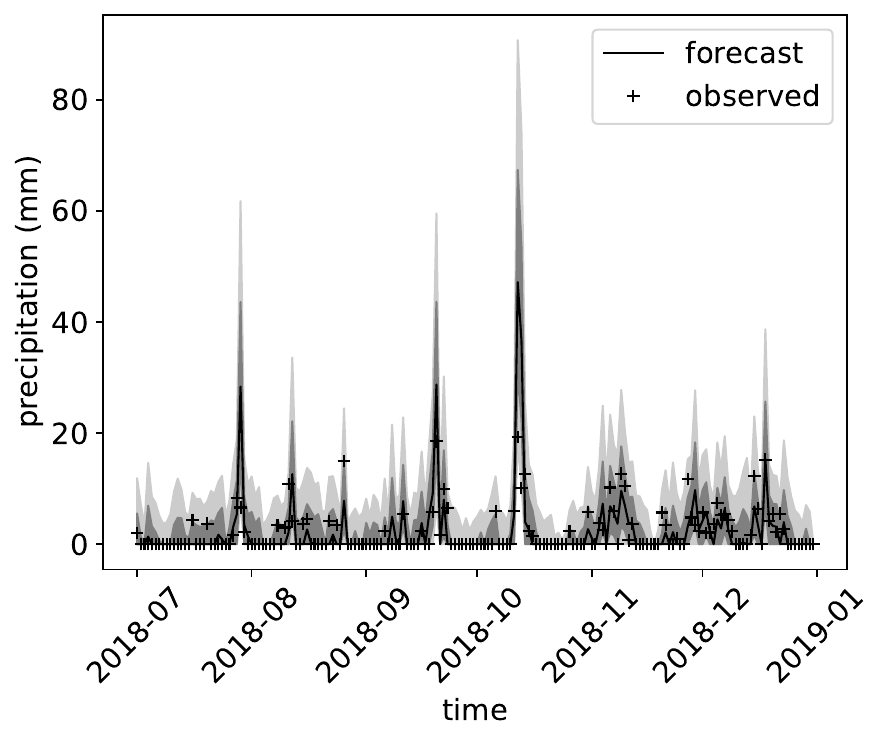}
    \caption{Year: 2018}
    \end{subfigure}
    \caption{\textbf{Cardiff: }The median (black solid line), 68\% (dark shading) and 95\% (light shading) highest density interval (HDI) of posterior predictive distribution of CP-MCMC and the observed precipitation at each of the days (black cross) for the last 6 months of the years (a) 2000 (b) 2006 (c) 2012 and (d) 2018.}
    \label{fig:cardiff_pred_obs}
\end{figure}

We train our model (called CP-MCMC in the following) at a single location close to Cardiff using datasets of length 1 year, 5 years, 10 years and 20 years up to 31st of December 1999 and compare the predicted precipitation with the observed precipitation and prediction provided by IFS for 20 years starting from 1st January 2000 up to 31st July 2019. In Figure~\ref{fig:cardiff_pred_obs}, we plot the median (black solid line), 68\% and 95\% highest density interval (HDI) (dark and light-shaded area correspondingly) of the posterior predictive distribution using the model trained on 5 years of data and the observed precipitation values at each of the days (black cross) for the last 6 months of the years 2000, 2006, 2012 and 2018. The last 6 months were chosen for the purpose of visualisation as it consists of both dry and wet days correspondingly in Summer and Winter seasons of Wales. For the black crosses exactly on the $x$-axis (\SI{0}{\milli\metre} of precipitation), we notice that the width of the dark shaded area corresponding to 68\% HDI almost goes to \SI{0}{\milli\metre} for those dry days. 
This indicates that our posterior predictive distribution of precipitation is a mixture between one distribution with all mass concentrated at \SI{0}{\milli\metre} and one distribution that is continuous over non-zero precipitation values with weights proportional to the probability of a dry day and a wet day. 
We also notice that a high proportion of the observed precipitation values for wet days also fall inside the HDIs predicted by our model, validating our model.

\begin{figure}[ht!]
\centering
\begin{subfigure}{0.49\textwidth}
    \includegraphics[width=\textwidth]{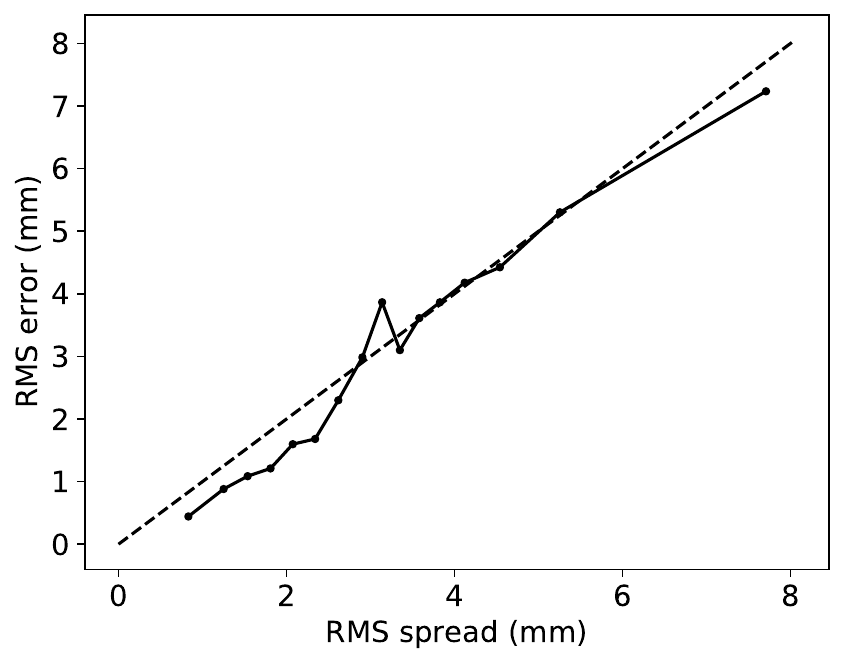}
    \caption{Spread-skill}
    \label{fig:spread_skill_cardiff}
\end{subfigure}
\begin{subfigure}{0.49\textwidth}
    \includegraphics[width=\textwidth]{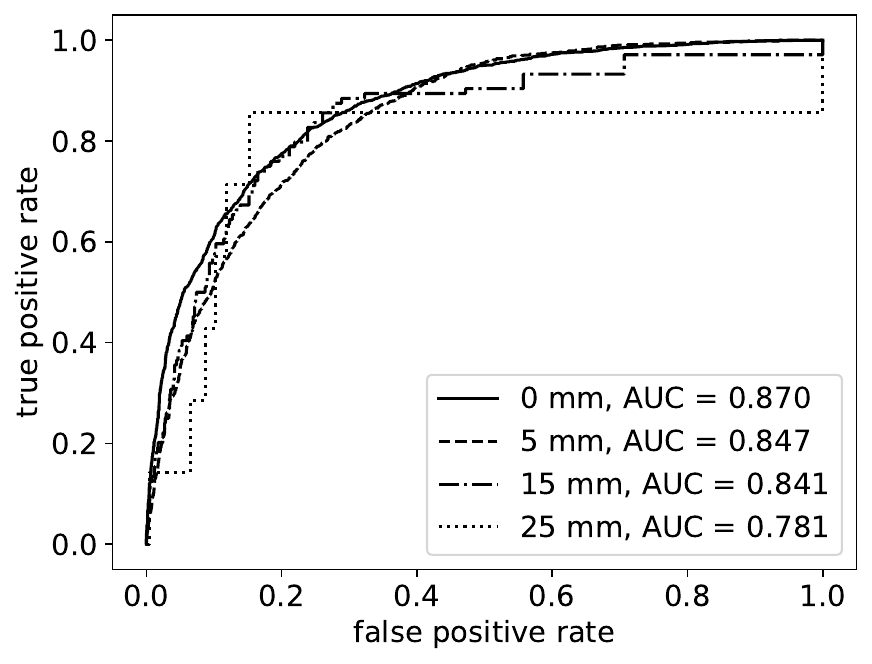}
    \caption{ROC curve.}
    \label{fig:cardiff_single_roc}
\end{subfigure}
\begin{subfigure}{0.49\textwidth}
    \includegraphics[width=\textwidth]{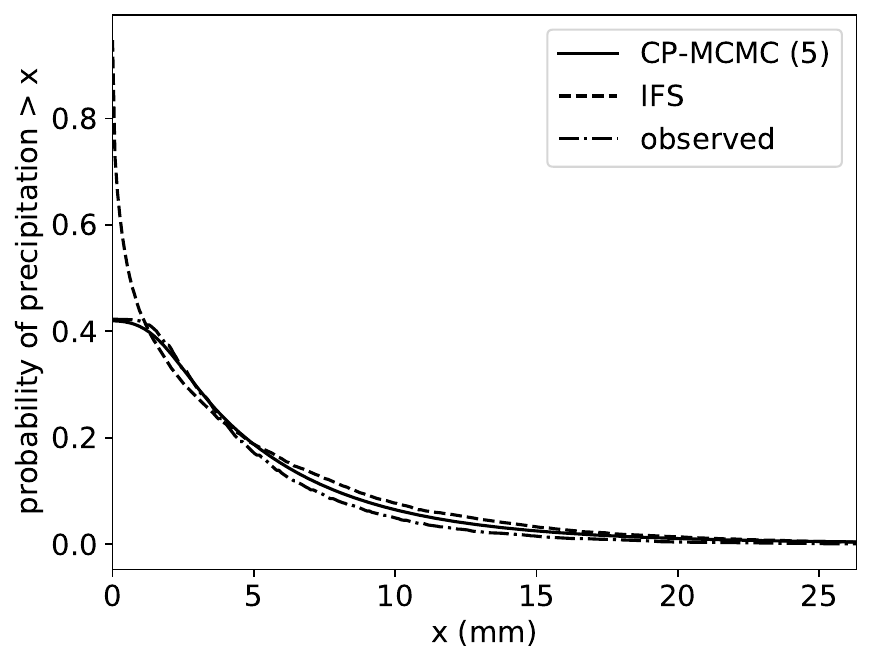}
    \caption{Probability of precipitation $>x$.}
    \label{fig:cardiff_surv}
\end{subfigure}
\caption{\textbf{Cardiff using CP-MCMC:} (a) Spread-skill relationship of our predictive distribution, plotting the RMS spread vs RMS error around the median of the predictive distribution of rainfall, (b) Receiver operating characteristic (ROC) curve plotting the true and false positive rates of detecting different levels of precipitation by CP-MCMC and the corresponding area under the curve (AUC) for precipitation prediction over the entire test, (c) Probability of precipitation $>x$ over the test set estimated by CP-MCMC, seen in observed data and predictions by IFS. CP-MCMC was trained on 5 years of data ending on 31st December 1999 precipitation was predicted over the test set (2000-2019).}
\label{fig:cardiff_roc_surv}
\end{figure}

The spread-skill graph is shown in Figure~\ref{fig:spread_skill_cardiff}, which allows evaluation of whether the model has correctly learnt state-dependence of the predictive uncertainty - whether it should be relatively confident or uncertain in any prediction. It shows the relationship between the RMS error and RMS spread \citep{leutbecher2008ensemble, christensen2015stochastic} in our forecast (may be referred to as the root mean square bias and the root mean variance in statistical literature). The RMS spread can be interpreted as the precision, how tight the credible intervals are in our forecast. The accuracy, or skill, is expressed in the RMS error, a lower RMS error correspond to higher accuracy. A forecast which is inaccurate should be reflected with larger uncertainty. This prevents forecasts which are confidently inaccurate or uncertainly correct. Given an ensemble of $M$ forecasts $\lbrace y^m_{t,s}: m = 1, \ldots, M \rbrace$, the squared error $b_{t,s}^2$ and squared spread $\sigma_{t,s}^2$, for a time point $t$ and location $s$, were calculated using
\begin{equation}
    b_{t,s}^2 = (\widehat{y}_{t,s} - y^{obs}_{t,s})^2, \ \    \sigma_{t,s}^2 = \dfrac{1}{M}\sum_{m=1}^M(y^m_{t,s} - \widehat{y}_{t,s})^2,
\end{equation}
where 
$y^{obs}_{t,s}$ and
$\widehat{y}_{t,s}$ stands correspondingly for observed precipitation and median of the ensemble forecast precipitation at time point $t$ and location $s$.
Calculating for $S$ locations and $T_\text{test}$ time points obtain a collection of $S \times T_\text{test}$ lots of $\{\sigma_{s,t}^2, b_{s,t}^2\}$ pairs. 
The days are grouped by forecast spread into 16 equally populated bins, and the RMS forecast spread and error relative to the forecast median are calculated for each bin and plotted in Figure~\ref{fig:spread_skill_cardiff}. In a model with the probabilistic component ideally represented, the observations are indistinguishable from individual predictions, and the RMS spread and error are identical in all bins. It can be seen from Figure~\ref{fig:spread_skill_cardiff} that the probabilistic component of CP-MCMC shows behaviour close to being ideal in this respect. This means it generally appropriately predicts low or high uncertainty when a low or high error can be expected, respectively.

Next, in Figure~\ref{fig:cardiff_single_roc}, we check the accuracy of estimates of the probability of an amount of rainfall on a day compared to the observed, by assessing the sensitivity of the detection of heavy precipitation over a certain amount, eg.~$>\SI{25}{\milli\metre}$. The prediction of a Boolean problem (eg.~is there heavy rain?) is known as testing in classical statistics or classification in machine learning. Given the ensemble of forecasts and for a particular point in space and time, a high proportion of samples which predict $>\SI{25}{\milli\metre}$ of precipitation should be an indication of heavy rain, this is known as a \textit{positive result}. We notice that the proportion of samples here indicating a heavy rain is subjective and will be considered as \textit{threshold} to be chosen. For a given \textit{threshold}, a test that correctly predicts heavy rain is known as a \textit{true positive}. But if the test predicts heavy rain on a day it did not occur, this is known as a \textit{false positive}. 
By treating precipitation at each day and point in space as separate independent events, a true positive rate (the proportion of points and days with heavy rain which was correctly detected) and the false positive rate (the proportion of points and days with no heavy rain with a positive result) can be obtained. Here, the true and false positive rate depends on the chosen threshold value. A receiver operating characteristic (ROC) curve plots the true and false positive rates as a parametric plot by varying the threshold. Different levels of precipitation can be tested, for example, $>\SI{5}{\milli\metre}$ for light rain, $>\SI{15}{\milli\metre}$ for medium rain  or $>\SI{25}{\milli\metre}$ for extreme events. For Wales in the test set (2000-2019), the proportion of days and locations with light, median and extreme precipitation observed are 22\%, 2.6\% and 0.41\% respectively.
The area under the ROC curve (AUC) can be used to assess how well the prediction for different levels of precipitation in the face of uncertainty was captured by the ensemble of forecasts \citep{metz1978basic, hanley1982meaning}. 

Furthermore, we test whether the estimated probability of a type of event (eg.~\text{precipitation}$>x$) over 20 years has matched with the observed probability, or whether they are at least similar. In Figure~\ref{fig:cardiff_surv}, we plot the estimated probability of our model, the predicted probability by IFS and the observed probability of  \text{precipitation}$>x$ by varying $x$ values, showing a much better agreement between our estimated probabilities and the observed probability when compared to the ones predicted by IFS specifically for events with low precipitation.

\begin{table}[ht!]
    \centering
    \begin{tabular}{|c|c|c|c|c|c|c|}
    \hline
         &Model & All year & Spring & Summer & Autumn & Winter  \\\hline
         \multirow{5}{*}{MAB}
         &CP-MCMC (1)   &      1.833 &   1.491 &   1.826 &   2.054 &   1.975 \\
         &CP-MCMC (5)  &      1.747 &   1.441 &   1.781 &   1.937 &   1.835 \\
         &CP-MCMC (10) &      1.716 &  1.414 &   1.732 &   1.894 &   1.828 \\
         &CP-MCMC (20) &      1.726 &  1.437 &   1.755 &   1.923 &   1.798 \\
         &IFS                 &      1.781 &   1.576 &   1.861 &   1.915 &   1.795 \\\hline
         \multirow{5}{*}{RMSB}
         &CP-MCMC (1)  &      3.670 &   2.971 &   3.728 &   4.303 &   3.595 \\
         &CP-MCMC (5)  &      3.475 &   2.923 &   3.683 &   3.862 &   3.369 \\
         &CP-MCMC (10) &      3.361 &   2.845 &   3.560 &   3.683 &   3.297 \\
         &CP-MCMC (20) &      3.428 &   2.962 &   3.607 &   3.775 &   3.315 \\
         &IFS          &      3.301 &   2.851 &   3.548 &   3.522 &   3.268 \\\hline
    \end{tabular}
    \caption{\textbf{Cardiff using CP-MCMC:} Mean absolute bias and root mean square bias, in \si{\milli\metre}, using the \emph{median} of the posterior predictive distribution by CP-MCMC over the test set (2000-2019) and the four seasons using different training sets (1, 5, 10 and 20 years ending at 31st December 1999) compared with the same for IFS.
    Key: `CP-MCMC (5)' stands for CP-MCMC model trained on 5 years of dataset up to 31st December 1999.}
    \label{tab:cardiff_mse_mab}
\end{table}

\begin{figure}[ht!]
\centering
        \begin{subfigure}{0.49\textwidth}
         \includegraphics[width= \textwidth]{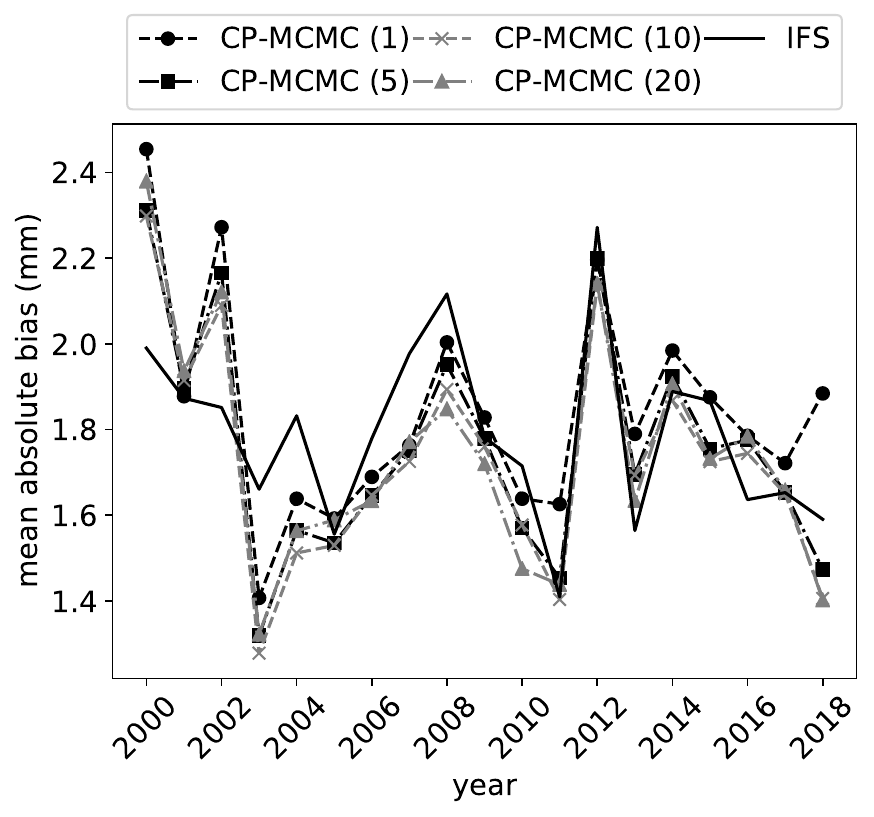}
    \caption{Mean absolute bias}
    \end{subfigure}
    \begin{subfigure}{0.49\textwidth}
         \includegraphics[width= \textwidth]{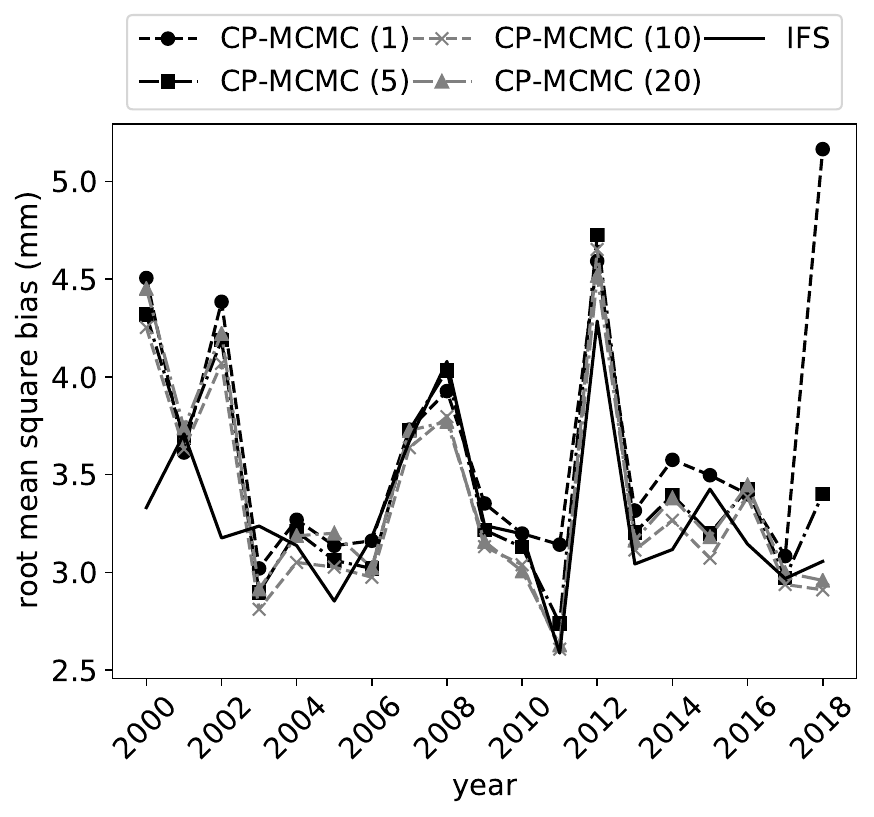}
    \caption{Root mean square bias}
    \end{subfigure}
    \caption{\textbf{Cardiff using CP-MCMC:} Mean absolute bias and root mean square bias using the \emph{median} of the posterior predictive distribution by CP-MCMC using different size training sets (1, 5, 10, 20 years ending at 31st December 1999) compared with IFS for 20 years upto 2019. Key: `CP-MCMC (5)' stands for CP-MCMC model trained on 5 years of dataset up to 31st December 1999.}
    \label{fig:cardiff_mse_mab}
\end{figure}

Although our model provides a probabilistic prediction of precipitation, we also compute a deterministic prediction for each day considering a central tendency eg.~median of the distribution of predicted rainfall for that day. In Table~\ref{tab:cardiff_mse_mab}  and Figure~\ref{fig:cardiff_mse_mab}, we compare the root mean squared bias (RMSB) and mean absolute bias (MAB) of the median prediction of CP-MCMC (trained on datasets of length correspondingly 1 year, 5 years, 10 years and 20 years) with the same for the daily prediction of IFS, showing a similar performance of our model with IFS. 
The RMSB and MAB over $S$ locations and $T$ time points are defined as 
$$ \text{RMSB} = \sqrt{\frac{1}{S\times T}\sum_{s=1}^S\sum_{t=1}^T ( y^{obs}_{t,s} - \widehat{y}_{t,s})^2}, \ \  \text{MAB} = \frac{1}{S \times T}\sum_{s=1}^S\sum_{t=1}^T |y^{obs}_{t,s} - \widehat{y}_{t,s}|$$
where $y^{obs}_{t,s}$ and $\widehat{y}_{t,s}$ stands correspondingly for observed precipitation and forecast precipitation by a methodology at time point $t$ and location $s$.
Extending the training data from one to five years improves prediction performance
but we do not get any further significant improvement when the model was trained using 10 and 20 years of data. Based on these findings, we train our model only using 5 years of data when predicting precipitation for Wales.

\begin{figure}[ht!]
    \centering
        \begin{subfigure}{0.49\textwidth}
         \includegraphics[width=\textwidth]{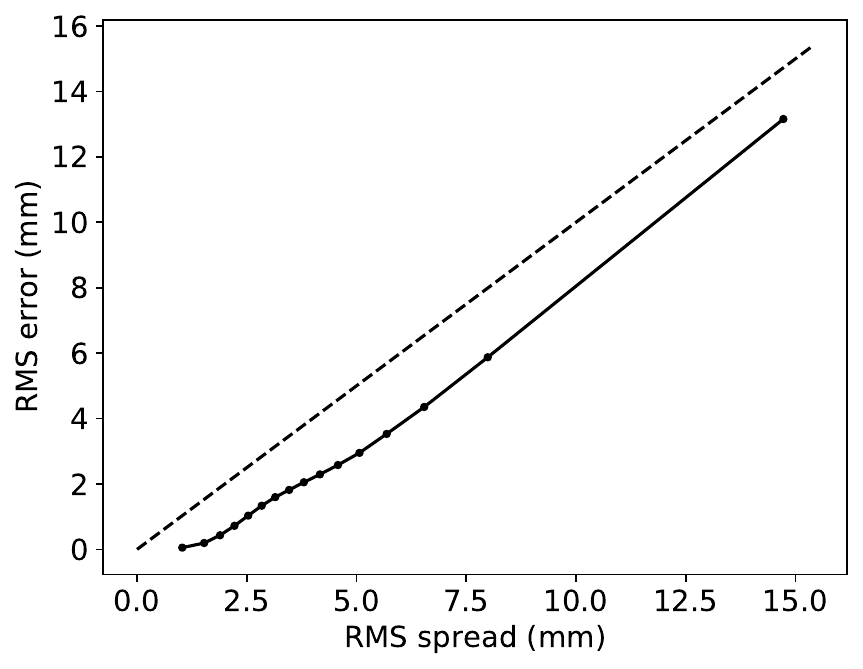}
          \caption{Skill-spread}
         \label{fig:spread_skill_wales}
    \end{subfigure}
    \begin{subfigure}{0.49\textwidth}
         \includegraphics[width=\textwidth]{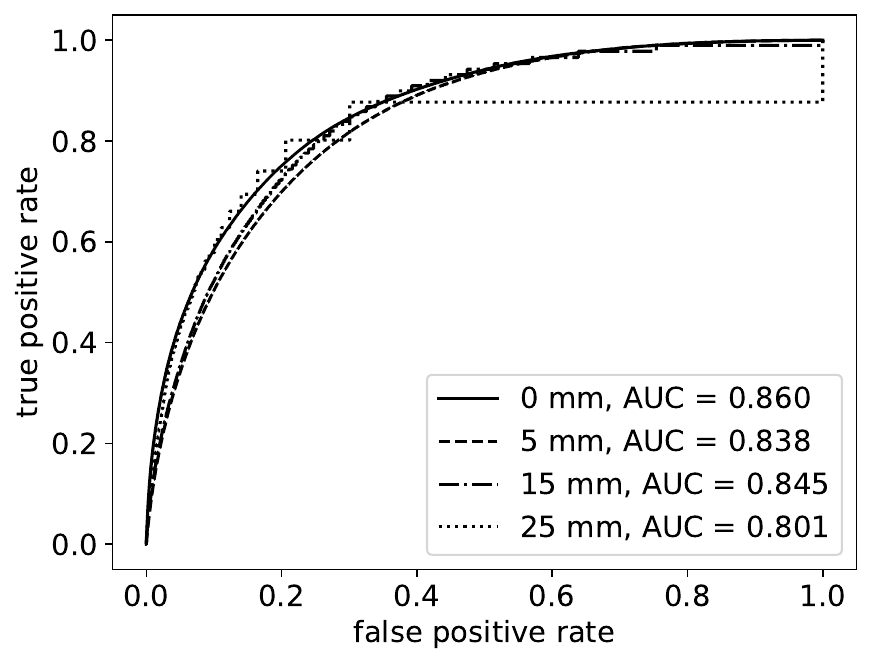}
         \caption{ROC curve}
         \label{fig:wales_roc_auc}
    \end{subfigure}
    \begin{subfigure}{0.49\textwidth}
     \includegraphics[width=\textwidth]{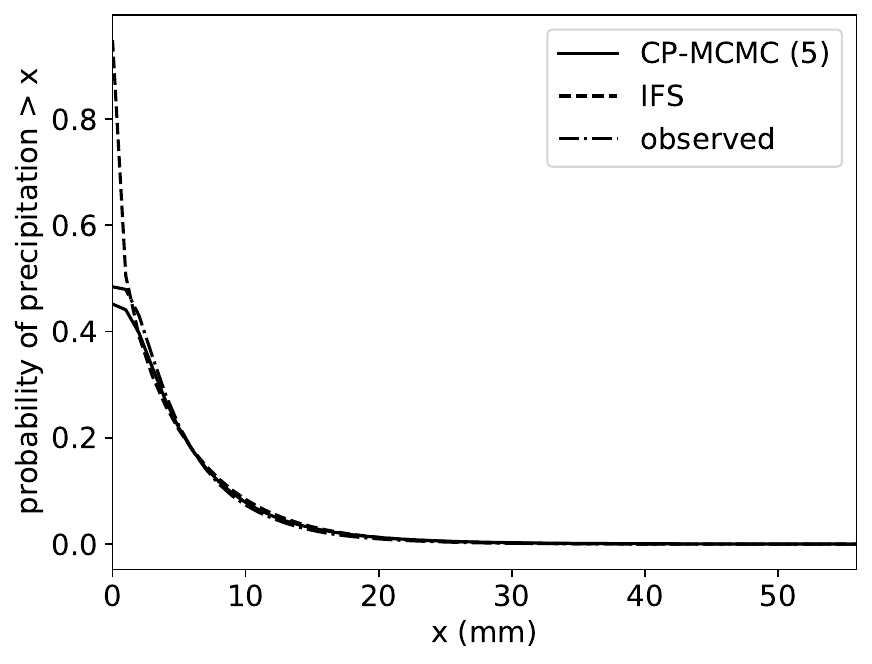}
          \caption{Probability of precipitation $>x$}
      \label{fig:wales_agg_compare}
    \end{subfigure}
     \begin{subfigure}{0.49\textwidth}
             \includegraphics[width=\textwidth]{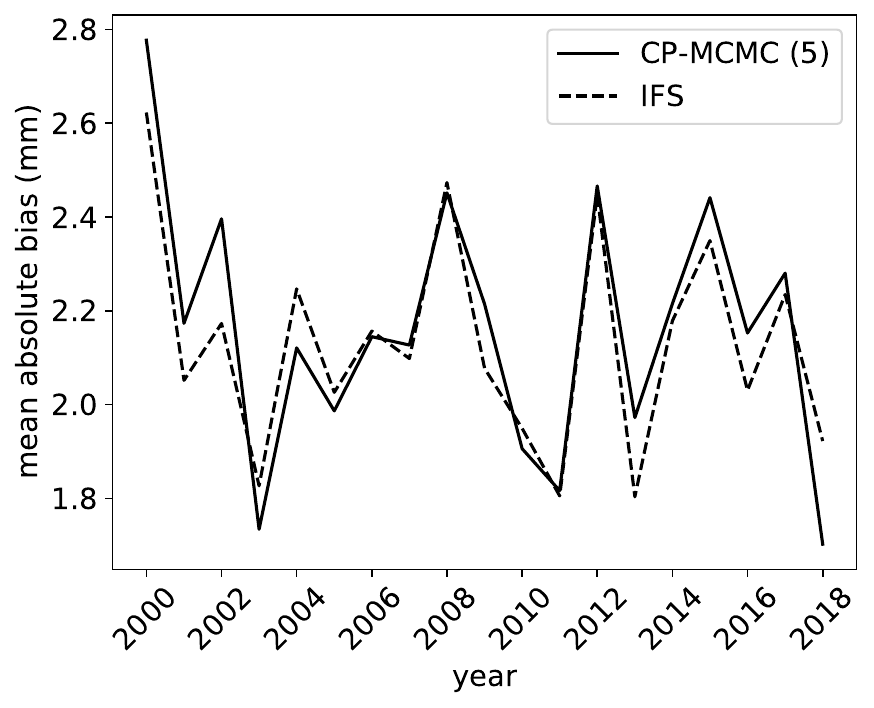}
             \caption{Temporal trend of MAB}
        \end{subfigure}
    \caption{\textbf{Wales using CP-MCMC:} (a) Spread-skill relationship of our predictive distribution using CP-MCMC, plotting the RMS spread vs RMS error around the median of the predictive distribution of rainfall, (b) Receiver operating characteristic (ROC) curve plotting the true and false positive rates of detecting different levels of precipitation by CP-MCMC and the corresponding area under the curve (AUC), (c) Probability of precipitation $>x$ over the test set estimated by CP-MCMC, seen in observed data and predictions by IFS, (d) Temporal trend of mean absolute bias of the \emph{median} of the posterior predictive distribution of CP-MCMC and IFS prediction over the test set (2000-2019), plotted in \si{\milli\metre}. CP-MCMC was trained on 5 years of data ending on 31st December 1999 precipitation was predicted over the test set (2000-2019).}
\end{figure}

\subsection{Rainfall prediction derived for  many independent grid points on Wales}
\label{sec:wales_result}
After validating our model for a single location, we train our model at 305 spatial points on Wales corresponding to the fine resolution grid. We train one model per grid point for the 5 years ending on 31st December 1999 and compare our probabilistic precipitation prediction with the observed precipitation and prediction of IFS over the nearly 20 years starting 1st of January 2000 and ending on 31st of July 2019. We decided to train our model from 5 years of data as this appears to be the optimal trade-off between computational complexity and prediction performance. 

The spread skill graph in Figure \ref{fig:spread_skill_wales} shows that our RMS spread is slightly too big and we could afford to decrease our credible intervals for more precise forecasts.
We assess the accuracy of probabilities estimated by our model for different events with different amounts of precipitation based on the ROC curve and AUC in Figure~\ref{fig:wales_roc_auc};
and by comparing the estimated probability and the observed probability of different precipitation events aggregated over the 20 years in Figure~\ref{fig:wales_agg_compare}. The figures show similar performance, to the single location inference for Cardiff in Section~\ref{sec:cardiff_result}.    

\subsection{Spatially coherent rainfall prediction over all of Wales} 

\begin{table}[ht!]
    \centering
    \begin{tabular}{|c|c|c|c|c|c|c|}\hline
         Error&Model & All year & Spring & Summer & Autumn & Winter  \\\hline
         \multirow{2}{*}{MAB}
         &CP-MCMC-GP (5) &   2.138 &   1.662 &   2.025 &   2.537 &   2.344 \\
         &CP-MCMC (5) &      2.145 &   1.667 &   2.033 &   2.546 &   2.353 \\
         &IFS         &      2.118 &   1.769 &   2.114 &   2.381 &   2.236 \\\hline
         \multirow{2}{*}{RMSB}
         &CP-MCMC-GP (5) &      4.115 &   3.224 &   4.007 &   4.848 &   4.247 \\
         &CP-MCMC (5) &      4.133 &   3.231 &   4.022 &   4.868 &   4.277 \\
         &IFS         &      3.756 &   3.084 &   3.756 &   4.262 &   3.878 \\\hline
    \end{tabular}
    \caption{\textbf{Wales using CP-MCMC-GP and CP-MCMC:} Mean absolute bias and root mean square bias, in \si{\milli\metre}, using the \emph{median} of the posterior predictive distribution by CP-MCMC-GP and CP-MCMC over the test set (2000-2019) and the four seasons at the same time period compared with the same for IFS.}
    \label{tab:wales_mse_mab}
\end{table} 

\begin{figure}[ht!]
    \centering
    \begin{subfigure}{0.99\textwidth}
        \centering
        \includegraphics[width=0.32\textwidth]{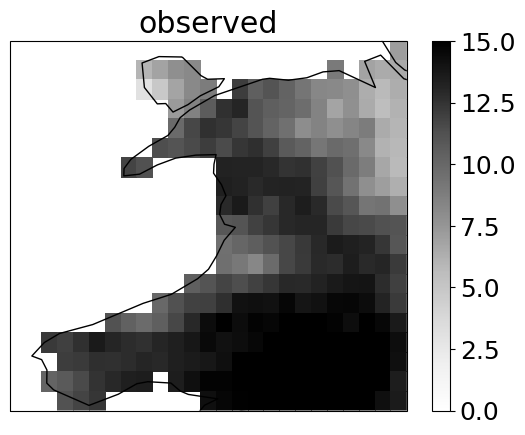}
        \includegraphics[width=0.32\textwidth]{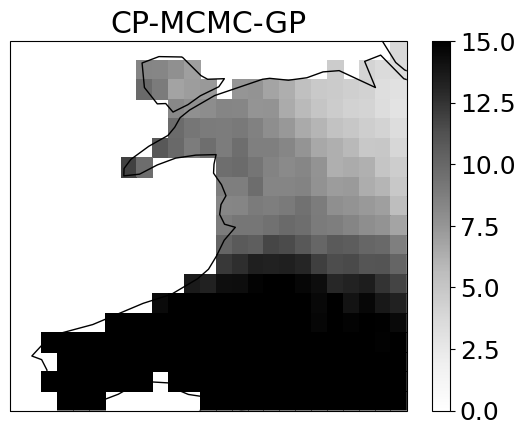}
        \includegraphics[width=0.32\textwidth]{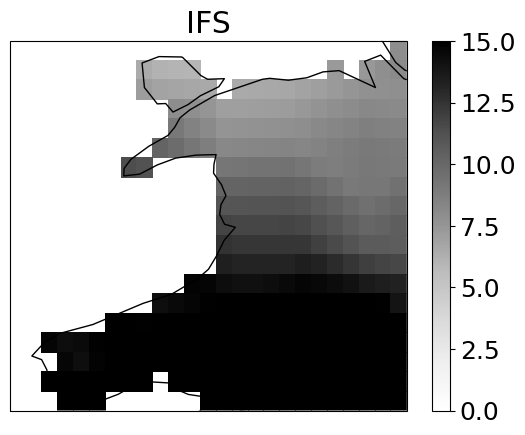}
        \caption{2000-12-31}
    \end{subfigure}
    \begin{subfigure}{0.99\textwidth}
        \centering
        \includegraphics[width=0.32\textwidth]{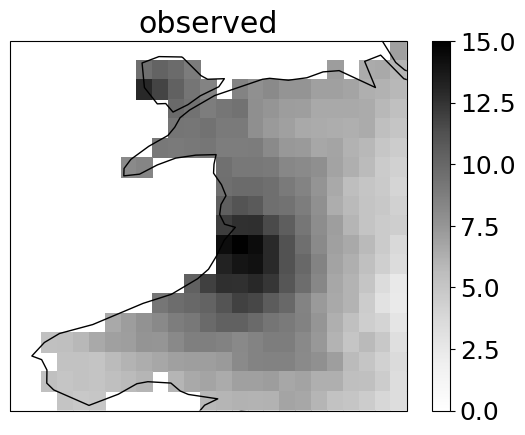}
        \includegraphics[width=0.32\textwidth]{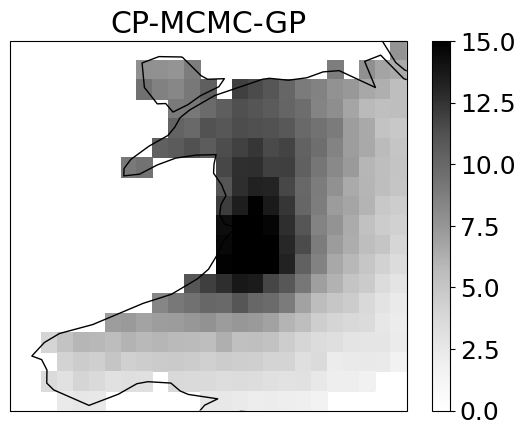}
        \includegraphics[width=0.32\textwidth]{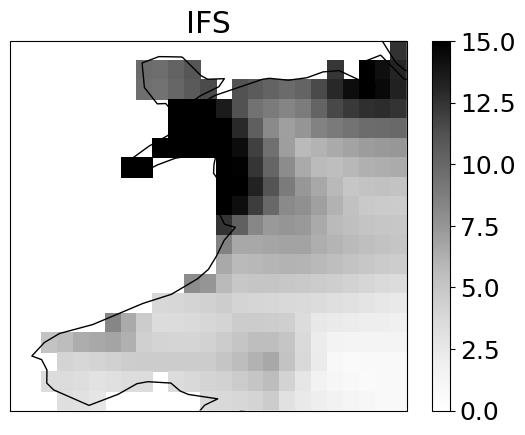}
        \caption{2006-08-31}
    \end{subfigure}
    \begin{subfigure}{0.99\textwidth}
        \centering
        \includegraphics[width=0.32\textwidth]{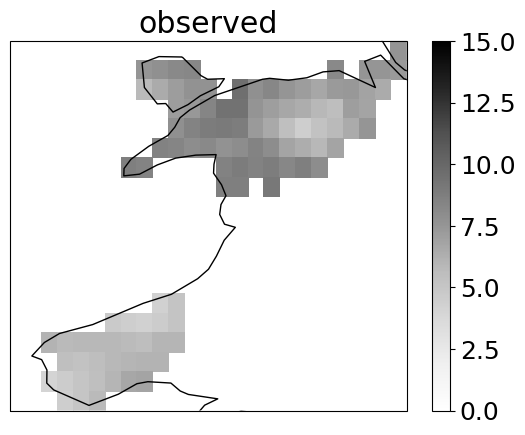}
        \includegraphics[width=0.32\textwidth]{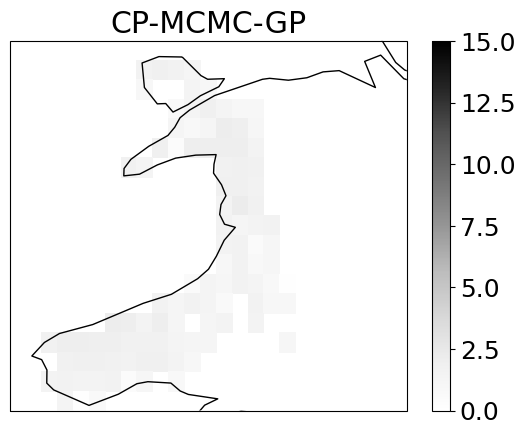}
        \includegraphics[width=0.32\textwidth]{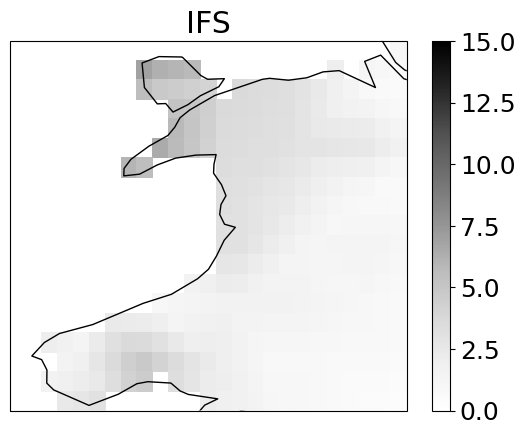}
        \caption{2012-11-18}
    \end{subfigure}
    \begin{subfigure}{0.99\textwidth}
        \centering
        \includegraphics[width=0.32\textwidth]{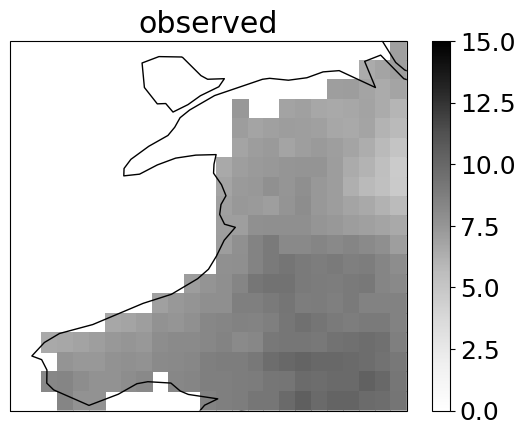}
        \includegraphics[width=0.32\textwidth]{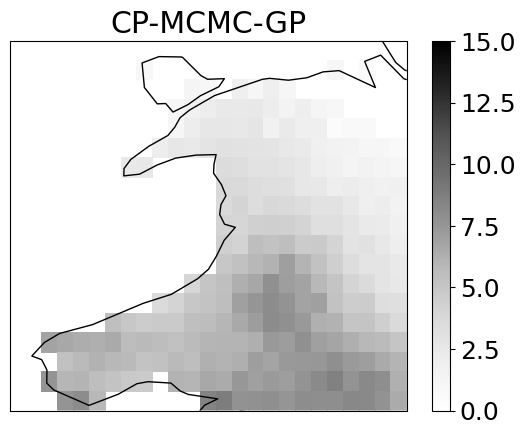}
        \includegraphics[width=0.32\textwidth]{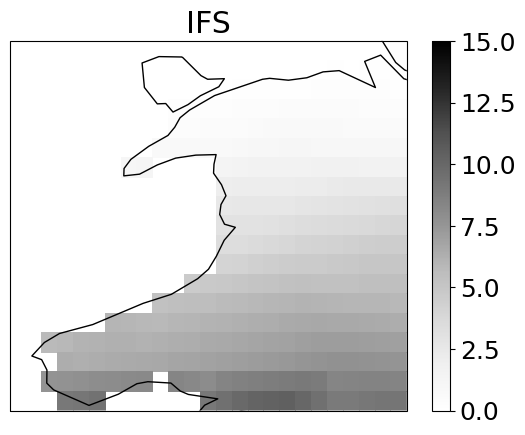}
        \caption{2018-09-22}
    \end{subfigure}
    
    \caption{\textbf{Wales using CP-MCMC-GP: } Median prediction of precipitation in \si{\milli\metre} by CP-MCMC-GP and IFS precipitation forecast compared to the observed.}
    \label{fig:wales_day_rainfall}
\end{figure}

In this Section, we produce rain fall prediction using CP-MCMC-GP as described in Section~\ref{sec:cpmcmcgp}, to achieve spatial coherence of the median prediction which is for example important to provide consistent results for flood predictions. In Figure~\ref{fig:wales_day_rainfall}, we illustrate the median prediction of precipitation by CP-MCMC-GP on 4 randomly chosen days with different amounts of rainfall from the autumn and winter seasons of 2000, 2006, 2012 and 2018 in comparison to the IFS forecast and the observed precipitation over Wales. This shows that our model prediction is able to pick up local variabilities in precipitation and have comparable to if not better performance than the IFS forecast.

\begin{figure}[ht!]
    \centering
        \begin{subfigure}{0.49\textwidth}
         \includegraphics[width=\textwidth]{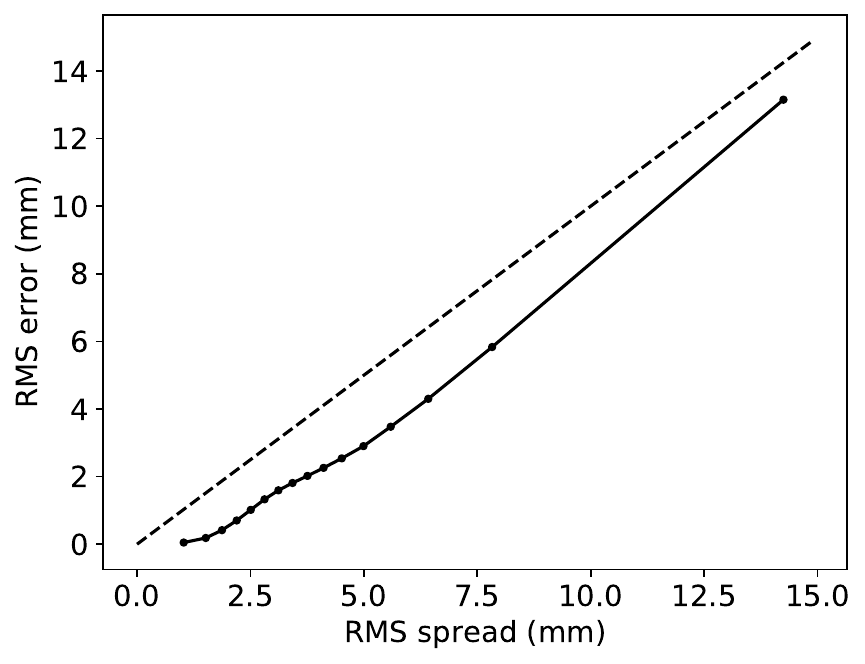}
          \caption{Skill-spread}
         \label{fig:spread_skill_wales_gp}
    \end{subfigure}
    \begin{subfigure}{0.49\textwidth}
         \includegraphics[width=\textwidth]{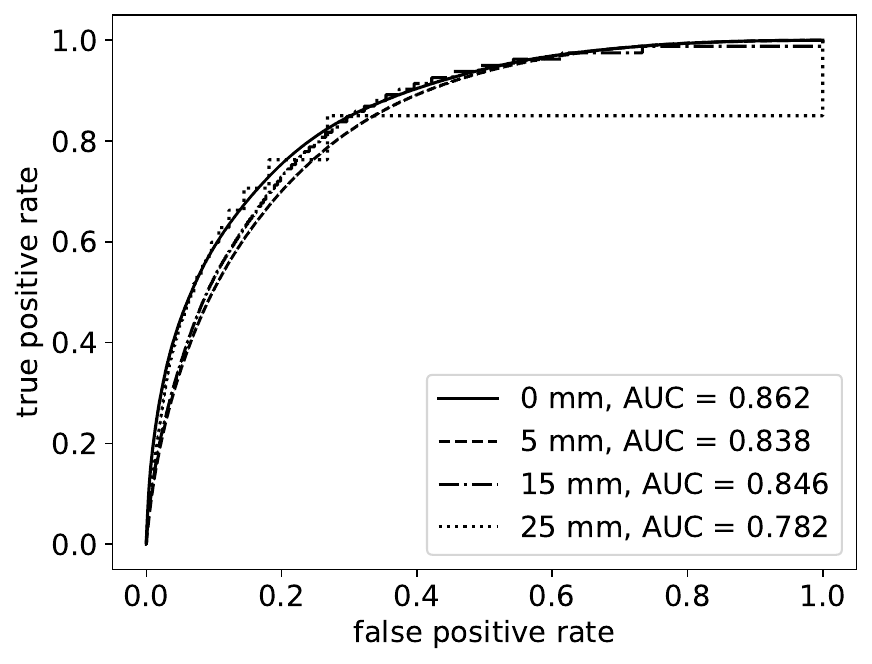}
         \caption{ROC curve}
         \label{fig:wales_roc_auc_gp}
    \end{subfigure}
    \begin{subfigure}{0.49\textwidth}
     \includegraphics[width=\textwidth]{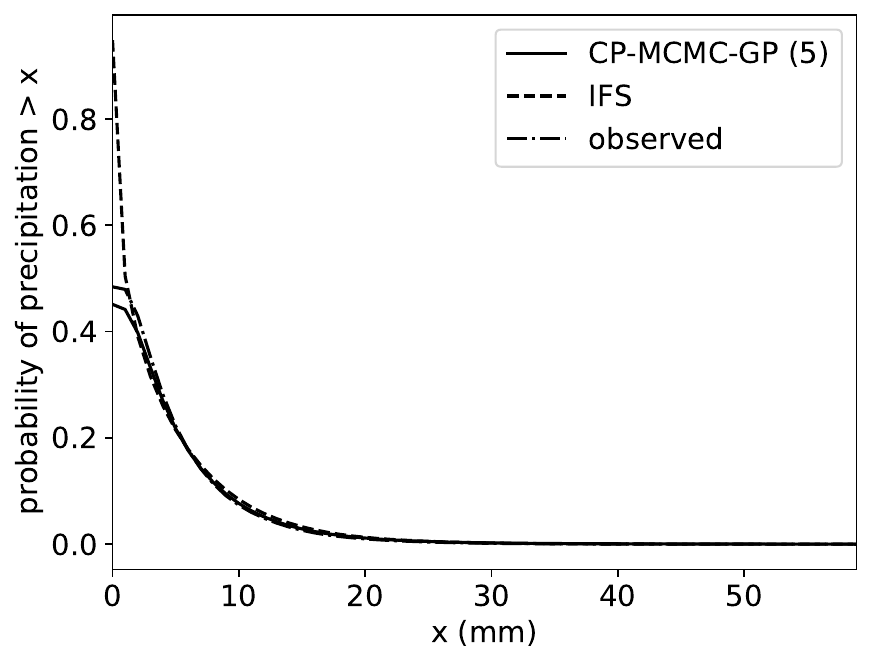}
          \caption{Probability of precipitation $>x$}
      \label{fig:wales_agg_compare_gp}
    \end{subfigure}
      \begin{subfigure}{0.49\textwidth}
             \includegraphics[width=\textwidth]{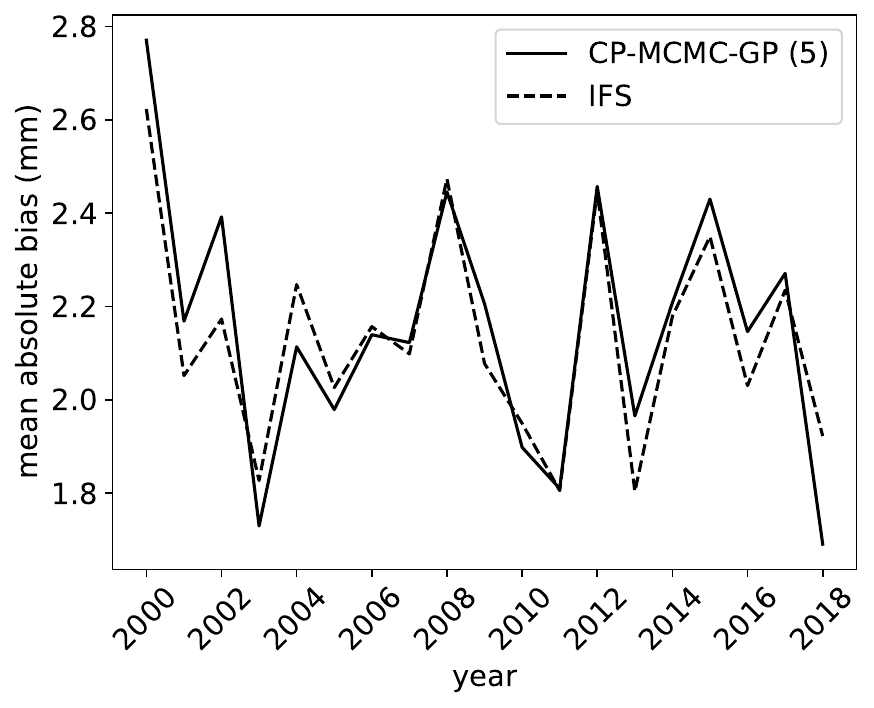}
             \caption{Temporal trend of MAB}
            \label{fig:wales_temp_mab_gp}
        \end{subfigure}
    \caption{\textbf{Wales using CP-MCMC-GP:} (a) Spread-skill relationship of our predictive distribution using CP-MCMC-GP, plotting the RMS spread vs RMS error around the median of the predictive distribution of rainfall, (b) Receiver operating characteristic (ROC) curve plotting the true and false positive rates of detecting different levels of precipitation by CP-MCMC-GP and the corresponding area under the curve (AUC), (c)     Probability of precipitation $>x$ over the test set estimated by CP-MCMC-GP, seen in observed data and predictions by IFS, (d) Temporal trend of mean absolute bias of the \emph{median} of the posterior predictive distribution of CP-MCMC-GP and IFS prediction over the test set (2000-2019), plotted in \si{\milli\metre}. CP-MCMC-GP was trained on 5 years of data ending on 31st December 1999 precipitation was predicted over the test set (2000-2019).}
       \label{fig:wales_comparison_cpmcmcgp}
\end{figure}

To asses whether our model can capture spatially coherent dependencies in median rainfall predictions, we compute the instantaneous cross-correlation of the median precipitation prediction at each location in Wales and the median precipitation prediction at the centre of Wales. The cross-correlation of our median forecast shows similarities to the observed precipitation. In Figure~\ref{fig:wales_comparison_cpmcmcgp}, we also provide the spread skill graph, the ROC curve, the comparison between the estimated probability and the observed probability of different precipitation events aggregated over the 20 years and the temporal trend of MAB compared to IFS, when the prediction is made using our CP-MCMC-GP model. 

\begin{figure}[ht!]
    \centering
    \begin{subfigure}{0.32\textwidth}
        \includegraphics[width=\textwidth]{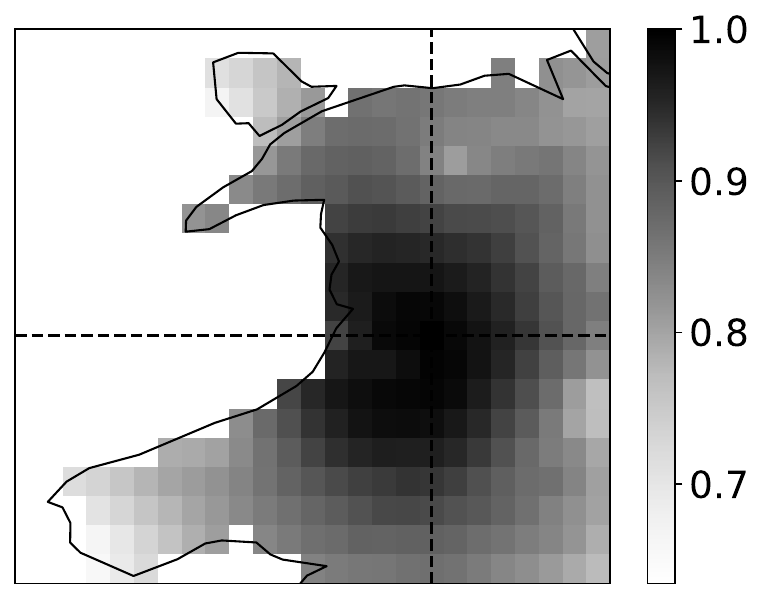}
        \caption{Observed}
    \end{subfigure}
    \begin{subfigure}{0.32\textwidth}
        \includegraphics[width=\textwidth]{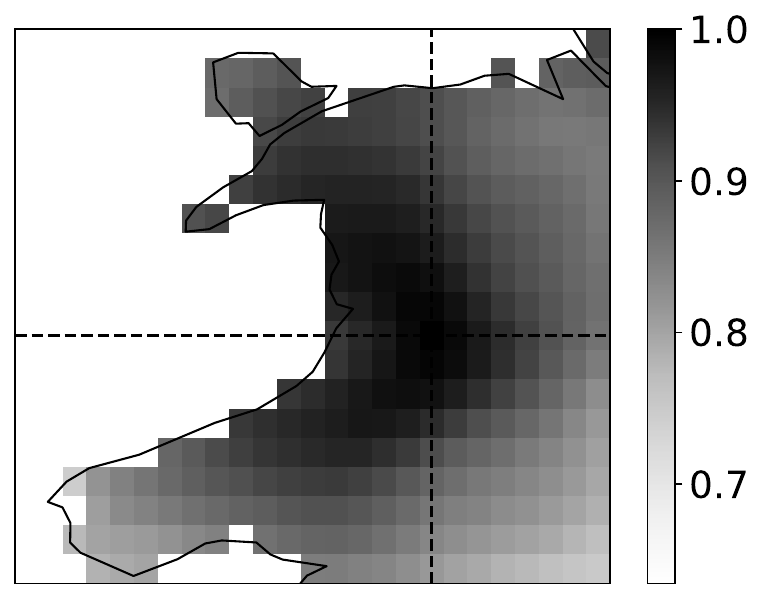}
    \caption{CP-MCMC-GP}
    \end{subfigure}
    \begin{subfigure}{0.32\textwidth}
        \includegraphics[width=\textwidth]{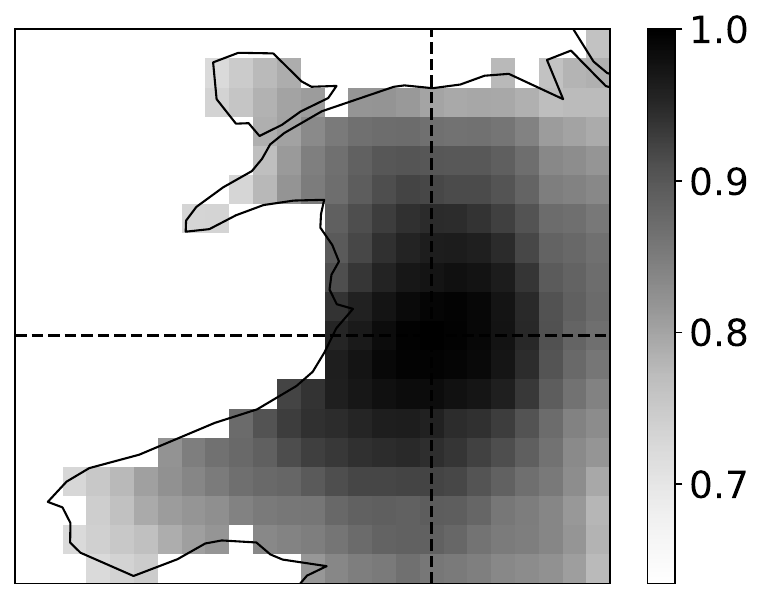}
    \caption{IFS}
    \end{subfigure}
    \caption{\textbf{Wales using CP-MCMC-GP: }Cross-correlation, with lag 0, of the precipitation between every pair of locations with the centre of mass of Wales using the observed (a), the median forecast by CP-MCMC-GP (b) and the IFS (c) predictions. The centre of mass is indicated by where the dashed lines intersect.}
    \label{fig:crosscorr_wales}
\end{figure}

\begin{figure}[ht!]
    \centering
        \begin{subfigure}{0.32\textwidth}
             \includegraphics[width=\textwidth]{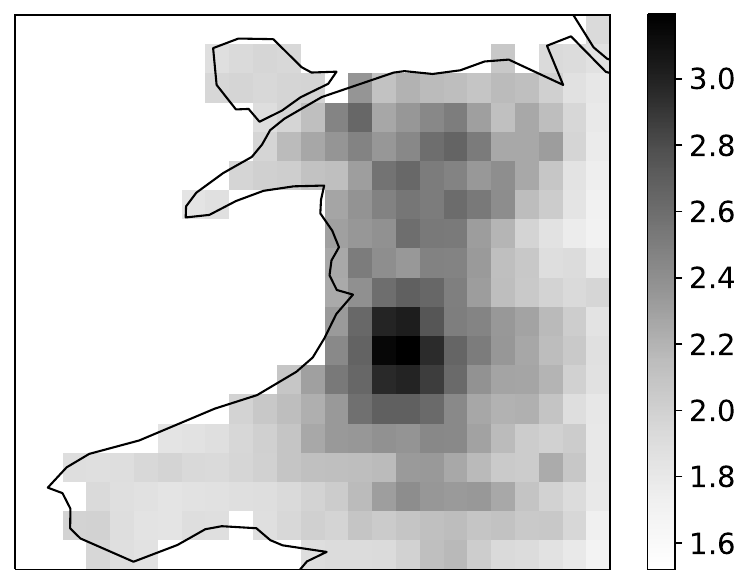}
             \caption{CP-MCMC-GP}
        \end{subfigure}
    \begin{subfigure}{0.32\textwidth}
         \includegraphics[width=\textwidth]{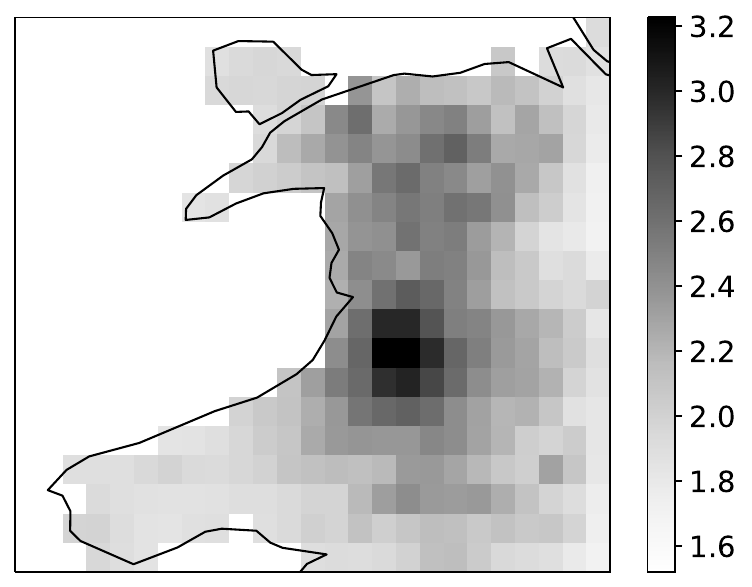}
         \caption{CP-MCMC}
    \end{subfigure}
    \begin{subfigure}{0.32\textwidth}
          \includegraphics[width=\textwidth]{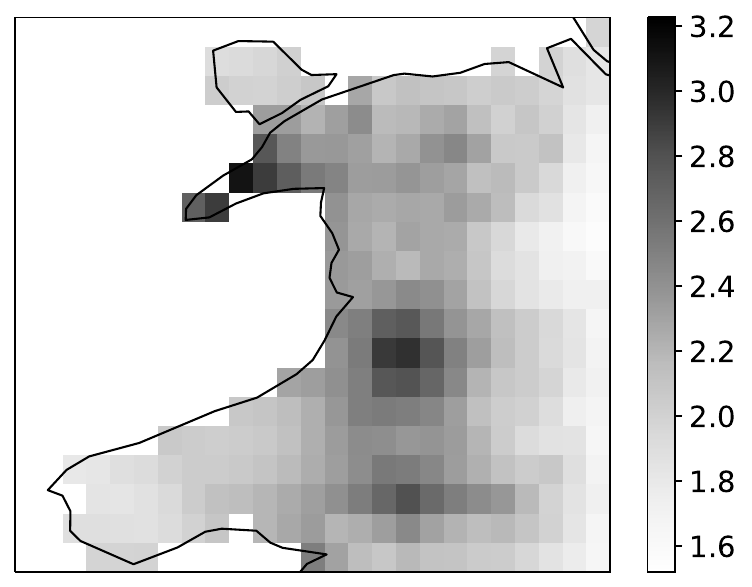}
          \caption{IFS}
    \end{subfigure}
    \caption{\textbf{Spatial trend of mean absolute bias: }of the \emph{median} of the posterior predictive distribution of CP-MCMC-GP, CP-MCMC and IFS prediction over the test set (2000-2019), plotted in \si{\milli\metre}.}
    \label{fig:wales_mse_mab}
\end{figure}

We compare the RMSB and MAB for the median prediction of CP-MCMC, CP-MCMC-GP and IFS over all years and different seasons in Table~\ref{tab:wales_mse_mab} and the spatial trend of MAB in Figure~\ref{fig:wales_mse_mab}. We can see that our median prediction performs similarly to IFS over the years and worse over the mountainous regions of Wales. Further, we notice improved performance for CP-MCMC-GP compared to CP-MCMC model in addition to providing spatial coherence.

\section{Conclusion and Future work}
\label{sec:conclusion}

In this paper, we proposed a statistical model of precipitation which takes the output of weather or climate models as input at a coarse resolution ($\sim\SI{50}{\kilo\metre}$) and use a Bayesian inferential framework to do a proper uncertainty quantification of the prediction of local precipitation at a high-resolution ($\sim\SI{10}{\kilo\metre}$). We first validate our model for a single location close to Cardiff and then illustrate its performance by producing predictions for Wales between 1st January 2000 and 31st July 2019, inclusive.
Regarding deterministic scores, our model is able to meet the quality of precipitation predictions of a full, three-dimensional atmospheric model (IFS) in short-term forecasts. Additionally, our tool is able to provide a reasonable representation of probabilistic forecast skill. This could have valuable practical applications for e.g.~flood modelling, where high-resolution precipitation predictions with realistic spatial and temporal variability are needed. 

Skill at predicting high precipitation events of our model can be improved by modelling the tail of the distribution of precipitation each day via an extreme value distribution \citep{behrens2004bayesian, macdonald2011flexible, ding2019modeling}. 
Further improvement in prediction can be achieved via imposing non-linear dependence of the parameters of compound Poisson distribution on the input variables, compared to the linear dependence considered here.   
A larger stencil of grid points for model input could also provide improvements to our models, observed in a second study of the same dataset that was based on the use of deep neural networks and was able to achieve even better results for deterministic predictions when taking also input vectors for neighbourhood locations into account \citep{adewoyin2020tru}. However, the deep learning study could not provide reasonable representations of forecast uncertainty. Non-linear dependence on the input variables including their values on the neighbouring locations may be the necessity while modelling rainfall at more complex topological areas. Results with our model could potentially also be improved by using a dynamic model selection approach to choose the number of ARMA terms $(p,q)$ in our model for each location $s$ using a dynamic global-local shrinkage prior \citep{bhadra2019lasso, kowal2019dynamic}. 

Some other directions of future work could consider more variables as input including topographic information at each location, using the wind velocity to direct the `direction' of the spatial dependency and a graphical model for the Poisson-element can add a spatial dependency \citep{yang2013poisson} for the hidden variables in our model. It should also be tested whether the model can be applied to more diverse training data with model data and observations distributed over the globe and whether it can be applied successfully for global post-processing of precipitation. Finally, feedback from experienced weather forecasters regarding the performance of our tool for specific weather situations would help to further understand the strength and weaknesses of the tool. 

\section{Acknowledgements}
The project was funded by Alan Turing Institute under Climate Action Pilot Projects Call.  We also acknowledge support from the HPC Facilities at the University of Warwick (Orac cluster). All of the code and data used in this project can be downloaded from \url{https://github.com/shermanlo77/cptimeseries}. We acknowledge the ERA5 and the E-OBS dataset from the Copernicus Climate Change Service, and the EU-FP6 project UERRA (\url{http://www.uerra.eu}) and the data providers in the ECA\&D project (\url{https://www.ecad.eu}). Peter Dueben gratefully acknowledges funding from the Royal Society for his University Research Fellowship and the
ESiWACE2 project which has received funding from
the European Union's Horizon 2020 research and
innovation programme under grant agreement No
823988. Peter Watson gratefully acknowledges funding (grant no.~NE/S014713/1) from the Natural Environment Research Council for his Independent Research Fellowship.

\bibliographystyle{apalike}

\begin{thebibliography}{}

\bibitem[Adewoyin et~al., 2020]{adewoyin2020tru}
Adewoyin, R., Dueben, P., Watson, P., He, Y., and Dutta, R. (2020).
\newblock Tru-net: A deep learning approach to high resolution prediction of
  rainfall.
\newblock {\em arXiv preprint arXiv:2008.09090}.

\bibitem[Behrens et~al., 2004]{behrens2004bayesian}
Behrens, C.~N., Lopes, H.~F., and Gamerman, D. (2004).
\newblock Bayesian analysis of extreme events with threshold estimation.
\newblock {\em Statistical Modelling}, 4(3):227--244.

\bibitem[Bertolacci et~al., 2019]{bertolacci2019climate}
Bertolacci, M., Cripps, E., Rosen, O., Lau, J.~W., Cripps, S., et~al. (2019).
\newblock Climate inference on daily rainfall across the {A}ustralian
  continent, 1876--2015.
\newblock {\em The Annals of Applied Statistics}, 13(2):683--712.

\bibitem[Bhadra et~al., 2019]{bhadra2019lasso}
Bhadra, A., Datta, J., Polson, N.~G., Willard, B., et~al. (2019).
\newblock Lasso meets horseshoe: A survey.
\newblock {\em Statistical Science}, 34(3):405--427.

\bibitem[Christensen et~al., 2015]{christensen2015stochastic}
Christensen, H.~M., Moroz, I., and Palmer, T. (2015).
\newblock Stochastic and perturbed parameter representations of model
  uncertainty in convection parameterization.
\newblock {\em Journal of the Atmospheric Sciences}, 72(6):2525--2544.

\bibitem[Cornes et~al., 2019]{E_obs}
Cornes, R., van~der Schrier, G., van~den Besselaar, E., and Jones, P. (2019).
\newblock An ensemble version of the e-obs temperature and precipitation
  datasets: version 21.0e.

\bibitem[Ding et~al., 2019]{ding2019modeling}
Ding, D., Zhang, M., Pan, X., Yang, M., and He, X. (2019).
\newblock Modeling extreme events in time series prediction.
\newblock In {\em Proceedings of the 25th ACM SIGKDD International Conference
  on Knowledge Discovery \& Data Mining}, pages 1114--1122.

\bibitem[Dueben and Bauer, 2018]{Dueben2018}
Dueben, P.~D. and Bauer, P. (2018).
\newblock Challenges and design choices for global weather and climate models
  based on machine learning.
\newblock {\em Geoscientific Model Development}, 11(10):3999--4009.

\bibitem[Dueben et~al., 2020]{Dueben2020}
Dueben, P.~D., Wedi, N., Saarinen, S., and Zeman, C. (2020).
\newblock Global simulations of the atmosphere at 1.45 km grid-spacing with the
  integrated forecasting system.
\newblock {\em Journal of the Meteorological Society of Japan. Ser. II},
  98(3):551--572.

\bibitem[Dunn, 2004]{dunn2004occurrence}
Dunn, P.~K. (2004).
\newblock Occurrence and quantity of precipitation can be modelled
  simultaneously.
\newblock {\em International Journal of Climatology: A Journal of the Royal
  Meteorological Society}, 24(10):1231--1239.

\bibitem[Dunn and Smyth, 2005]{dunn2005series}
Dunn, P.~K. and Smyth, G.~K. (2005).
\newblock Series evaluation of {T}weedie exponential dispersion model
  densities.
\newblock {\em Statistics and Computing}, 15(4):267--280.

\bibitem[Dzupire et~al., 2018]{dzupire2018poisson}
Dzupire, N.~C., Ngare, P., and Odongo, L. (2018).
\newblock A {P}oisson-gamma model for zero inflated rainfall data.
\newblock {\em Journal of Probability and Statistics}, 2018.

\bibitem[Fuhrer et~al., 2018]{Fuhrer2018}
Fuhrer, O., Chadha, T., Hoefler, T., Kwasniewski, G., Lapillonne, X.,
  Leutwyler, D., L\"uthi, D., Osuna, C., Sch\"ar, C., Schulthess, T.~C., and
  Vogt, H. (2018).
\newblock Near-global climate simulation at 1\,km resolution: establishing a
  performance baseline on 4888\,gpus with cosmo 5.0.
\newblock {\em Geoscientific Model Development}, 11(4):1665--1681.

\bibitem[Garc{\'\i}a et~al., 1995]{garcia1995analysis}
Garc{\'\i}a, J., Marroquin, A., Garrido, J., and Mateos, V. (1995).
\newblock Analysis of daily rainfall processes in lower extremadura ({S}pain)
  and homogenization of the data.
\newblock {\em Theoretical and Applied Climatology}, 51(1):75--87.

\bibitem[Geman and Geman, 1984]{geman1984stochastic}
Geman, S. and Geman, D. (1984).
\newblock Stochastic relaxation, {G}ibbs distributions, and the {B}ayesian
  restoration of images.
\newblock {\em Institute of Electrical and Electronics Engineers Transactions
  on Pattern Analysis and Machine Intelligence}, PAMI-6(6):721--741.

\bibitem[Geyer, 1992]{geyer1992practical}
Geyer, C.~J. (1992).
\newblock Practical {M}arkov chain {M}onte {C}arlo.
\newblock {\em Statistical Science}, pages 473--483.

\bibitem[Grönquist et~al., 2020]{Groenquist2020}
Grönquist, P., Yao, C., Ben-Nun, T., Dryden, N., Dueben, P., Li, S., and
  Hoefler, T. (2020).
\newblock Deep learning for post-processing ensemble weather forecasts.

\bibitem[Guti{\'e}rrez et~al., 2019]{gutierrez2019intercomparison}
Guti{\'e}rrez, J.~M., Maraun, D., Widmann, M., Huth, R., Hertig, E., Benestad,
  R., R{\"o}ssler, O., Wibig, J., Wilcke, R., Kotlarski, S., et~al. (2019).
\newblock An intercomparison of a large ensemble of statistical downscaling
  methods over europe: Results from the value perfect predictor
  cross-validation experiment.
\newblock {\em International journal of climatology}, 39(9):3750--3785.

\bibitem[Haario et~al., 2001]{haario2001adaptive}
Haario, H., Saksman, E., and Tamminen, J. (2001).
\newblock An adaptive {M}etropolis algorithm.
\newblock {\em Bernoulli}, 7(2):223--242.

\bibitem[Hanley and McNeil, 1982]{hanley1982meaning}
Hanley, J.~A. and McNeil, B.~J. (1982).
\newblock The meaning and use of the area under a receiver operating
  characteristic ({ROC}) curve.
\newblock {\em Radiology}, 143(1):29--36.

\bibitem[Hastings, 1970]{hastings1970monte}
Hastings, W.~K. (1970).
\newblock Monte {C}arlo sampling methods using {M}arkov chains and their
  applications.
\newblock {\em Biometrika}, 57(1):97--109.

\bibitem[Hersbach et~al., 2020]{ERA5}
Hersbach, H., Bell, B., Berrisford, P., Hirahara, S., Horányi, A., and
  Muñoz-Sabater, J. (2020).
\newblock The era5 global reanalysis.
\newblock {\em Quarterly Journal of the Royal Meteorological Society},
  146(730):1999--2049.

\bibitem[Hewson and Pillosu, 2020]{Hewson2020}
Hewson, T.~D. and Pillosu, F.~M. (2020).
\newblock A new low-cost technique improves weather forecasts across the world.

\bibitem[Hoffman and Gelman, 2014]{hoffman2014no}
Hoffman, M.~D. and Gelman, A. (2014).
\newblock The {N}o-{U}-turn sampler: {A}daptively setting path lengths in
  {H}amiltonian {M}onte {C}arlo.
\newblock {\em Journal of Machine Learning Research}, 15(1):1593--1623.

\bibitem[Holsclaw et~al., 2017]{holsclaw2017bayesian}
Holsclaw, T., Greene, A.~M., Robertson, A.~W., Smyth, P., et~al. (2017).
\newblock Bayesian nonhomogeneous markov models via p{\'o}lya-gamma data
  augmentation with applications to rainfall modeling.
\newblock {\em The Annals of Applied Statistics}, 11(1):393--426.

\bibitem[Kowal et~al., 2019]{kowal2019dynamic}
Kowal, D.~R., Matteson, D.~S., and Ruppert, D. (2019).
\newblock Dynamic shrinkage processes.
\newblock {\em Journal of the Royal Statistical Society: Series B (Statistical
  Methodology)}, 81(4):781--804.

\bibitem[LeCam, 1953]{lecam1953some}
LeCam, L. (1953).
\newblock On some asymptotic properties of maximum likelihood estimates and
  related bayes estimates.
\newblock {\em Univ. California Pub. Statist.}, 1:277--330.

\bibitem[Leutbecher and Palmer, 2008]{leutbecher2008ensemble}
Leutbecher, M. and Palmer, T.~N. (2008).
\newblock Ensemble forecasting.
\newblock {\em Journal of computational physics}, 227(7):3515--3539.

\bibitem[Lo, 2020]{lo2020characterisation}
Lo, S.~E. (2020).
\newblock {\em Characterisation of Computed Tomography Noise in Projection
  Space with Applications to Additive Manufacturing}.
\newblock PhD thesis, University of Warwick.

\bibitem[MacDonald et~al., 2011]{macdonald2011flexible}
MacDonald, A., Scarrott, C.~J., Lee, D., Darlow, B., Reale, M., and Russell, G.
  (2011).
\newblock A flexible extreme value mixture model.
\newblock {\em Computational Statistics \& Data Analysis}, 55(6):2137--2157.

\bibitem[Maraun et~al., 2019]{maraun2019statistical}
Maraun, D., Widmann, M., and Guti{\'e}rrez, J.~M. (2019).
\newblock Statistical downscaling skill under present climate conditions: A
  synthesis of the value perfect predictor experiment.
\newblock {\em International Journal of Climatology}, 39(9):3692--3703.

\bibitem[Metropolis et~al., 1953]{metropolis1953equation}
Metropolis, N., Rosenbluth, A.~W., Rosenbluth, M.~N., Teller, A.~H., and
  Teller, E. (1953).
\newblock Equation of state calculations by fast computing machines.
\newblock {\em The Journal of Chemical Physics}, 21(6):1087--1092.

\bibitem[Metz, 1978]{metz1978basic}
Metz, C.~E. (1978).
\newblock Basic principles of {ROC} analysis.
\newblock {\em Seminars in Nuclear Medicine}, 8(4):283--298.

\bibitem[Murphy et~al., 2018]{Murphy2018}
Murphy, J., Harris, G., Sexton, D., Kendon, E., Bett, P., and Clark, R. (2018).
\newblock Ukcp18 land projections: Science report.
\newblock Technical report, Met Office.

\bibitem[Murray et~al., 2010]{murray2010elliptical}
Murray, I., Adams, R.~P., and MacKay, D.~J. (2010).
\newblock Elliptical slice sampling.
\newblock In {\em Proceedings of the 13th International Conference on
  Artificial Intelligence and Statistics}.

\bibitem[Neal, 2003]{neal2003slice}
Neal, R.~M. (2003).
\newblock Slice sampling.
\newblock {\em The Annals of Statistics}, 31(3):705--741.

\bibitem[Neal, 2011]{neal2011mcmc}
Neal, R.~M. (2011).
\newblock {MCMC} using {H}amiltonian dynamics.
\newblock In Brooks, S., Gelman, A., Jones, G., and Meng, X.-L., editors, {\em
  Handbook of Markov Chain {M}onte {C}arlo}, chapter~5, pages 113--162. CRC
  press.

\bibitem[Neumann et~al., 2019]{Neumann2018}
Neumann, P., Dueben, P., Adamidis, P., Bauer, P., Brück, M., Kornblueh, L.,
  Klocke, D., Stevens, B., Wedi, N., and Biercamp, J. (2019).
\newblock Assessing the scales in numerical weather and climate predictions:
  will exascale be the rescue?
\newblock {\em Philosophical Transactions of the Royal Society A: Mathematical,
  Physical and Engineering Sciences}, 377(2142):20180148.

\bibitem[Rasp et~al., 2020]{Rasp2020}
Rasp, S., Dueben, P.~D., Scher, S., Weyn, J.~A., Mouatadid, S., and Thuerey, N.
  (2020).
\newblock Weatherbench: A benchmark dataset for data-driven weather
  forecasting.
\newblock {\em arXiv preprint arXiv:2002.00469}.

\bibitem[Rasp and Lerch, 2018]{Rasp2018}
Rasp, S. and Lerch, S. (2018).
\newblock {Neural Networks for Postprocessing Ensemble Weather Forecasts}.
\newblock {\em Monthly Weather Review}, 146(11):3885--3900.

\bibitem[Revfeim, 1984]{revfeim1984initial}
Revfeim, K. J.~A. (1984).
\newblock An initial model of the relationship between rainfall events and
  daily rainfalls.
\newblock {\em Journal of Hydrology}, 75(1):357 -- 364.

\bibitem[Robert and Casella, 2005]{Robert_2005}
Robert, C.~P. and Casella, G. (2005).
\newblock {\em {M}onte Carlo Statistical Methods (Springer Texts in
  Statistics)}.
\newblock Springer-Verlag New York, Inc., Secaucus, NJ, USA.

\bibitem[Roberts and Rosenthal, 2009]{roberts2009examples}
Roberts, G.~O. and Rosenthal, J.~S. (2009).
\newblock Examples of adaptive {MCMC}.
\newblock {\em Journal of Computational and Graphical Statistics},
  18(2):349--367.

\bibitem[Schaller et~al., 2016]{schaller2016human}
Schaller, N., Kay, A.~L., Lamb, R., Massey, N.~R., Van~Oldenborgh, G.~J., Otto,
  F.~E., Sparrow, S.~N., Vautard, R., Yiou, P., Ashpole, I., et~al. (2016).
\newblock Human influence on climate in the 2014 southern england winter floods
  and their impacts.
\newblock {\em Nature Climate Change}, 6(6):627.

\bibitem[Scher and Messori, 2019]{Scher2019}
Scher, S. and Messori, G. (2019).
\newblock Weather and climate forecasting with neural networks: using general
  circulation models (gcms) with different complexity as a study ground.
\newblock {\em Geoscientific Model Development}, 12(7):2797--2809.

\bibitem[{Schulthess} et~al., 2019]{Schulthess2019}
{Schulthess}, T.~C., {Bauer}, P., {Wedi}, N., {Fuhrer}, O., {Hoefler}, T., and
  {Schär}, C. (2019).
\newblock Reflecting on the goal and baseline for exascale computing: A roadmap
  based on weather and climate simulations.
\newblock {\em Computing in Science Engineering}, 21(1):30--41.

\bibitem[Vogel et~al., 2018]{Vogel2018}
Vogel, P., Knippertz, P., Fink, A.~H., Schlueter, A., and Gneiting, T. (2018).
\newblock {Skill of Global Raw and Postprocessed Ensemble Predictions of
  Rainfall over Northern Tropical Africa}.
\newblock {\em Weather and Forecasting}, 33(2):369--388.

\bibitem[Wallemacq, 2018]{wallemacq2018economic}
Wallemacq, P. (2018).
\newblock Economic losses, poverty and disasters 1998-2017.
\newblock DOI10.13140/rg.2.2.35610.08643.

\bibitem[Weyn et~al., 2019]{Weyn2019}
Weyn, J.~A., Durran, D.~R., and Caruana, R. (2019).
\newblock Can machines learn to predict weather? using deep learning to predict
  gridded 500-hpa geopotential height from historical weather data.
\newblock {\em Journal of Advances in Modeling Earth Systems},
  11(8):2680--2693.

\bibitem[Williams and Rasmussen, 1996]{williams1996gaussian}
Williams, C. K.~I. and Rasmussen, C.~E. (1996).
\newblock Gaussian processes for regression.
\newblock In {\em Proceedings of Advances in Neural Information Processing
  Systems}, pages 514--520.

\bibitem[{World Economic Forum}, 2019]{wef2019global}
{World Economic Forum} (2019).
\newblock volume https://www.weforum.org/reports/the-global-risks-report-2019.

\bibitem[Yang et~al., 2013]{yang2013poisson}
Yang, E., Ravikumar, P.~K., Allen, G.~I., and Liu, Z. (2013).
\newblock On {P}oisson graphical models.
\newblock In {\em Proceedings of Advances in Neural Information Processing
  Systems}, pages 1718--1726.

\end{thebibliography}

\end{document}